\documentclass[review,authoryear,12pt]{elsarticle}
\usepackage[hmargin=1in]{geometry} 
\usepackage{amsmath}

\journal{Icarus}

\begin{document}
\begin{frontmatter}

\title{GCM Simulations of Titan's Middle and Lower Atmosphere and Comparison to Observations}
\author{\textit{Accepted for publication in Icarus}\\ \vspace{10pt} 
Juan M. Lora$^{a,1}$, Jonathan I. Lunine$^b$, Joellen L. Russell$^a$}
\address{$^a$Department of Planetary Sciences, University of Arizona, Tucson, AZ 85721\\
$^b$Center for Radiophysics and Space Research, Cornell University, Ithaca, NY 14853\\
$^1$Present address: Department of Earth, Planetary, and Space Sciences, University of California, Los Angeles, CA 90095}

\begin{abstract}
Simulation results are presented from a new general circulation model (GCM) of Titan, the Titan Atmospheric Model (TAM), which couples the Flexible Modeling System (FMS) spectral dynamical core to a suite of external/sub-grid-scale physics. These include a new non-gray radiative transfer module that takes advantage of recent data from Cassini-Huygens, large-scale condensation and quasi-equilibrium moist convection schemes, a surface model with ``bucket" hydrology, and boundary layer turbulent diffusion. The model produces a realistic temperature structure from the surface to the lower mesosphere, including a stratopause, as well as satisfactory superrotation. The latter is shown to depend on the dynamical core's ability to build up angular momentum from surface torques. Simulated latitudinal temperature contrasts are adequate, compared to observations, and polar temperature anomalies agree with observations. In the lower atmosphere, the insolation distribution is shown to strongly impact turbulent fluxes, and surface heating is maximum at mid-latitudes. Surface liquids are unstable at mid- and low-latitudes, and quickly migrate poleward. The simulated humidity profile and distribution of surface temperatures, compared to observations, corroborate the prevalence of dry conditions at low latitudes. Polar cloud activity is well represented, though the observed mid-latitude clouds remain somewhat puzzling, and some formation alternatives are suggested.
\end{abstract}
\end{frontmatter}

\section{Introduction}
Observations of Titan since the time of the Voyager 1 flyby have prompted the development of several general circulation models (GCMs) to study its atmosphere. The first GCM of Titan \citep{Hourdin95} studied the development of atmospheric superrotation, showing relative agreement with then-current observations. Subsequent axisymmetric (two-dimensional) models provided a variety of additional insights into Titan's climate processes, including the first studies of the methane and ethane hydrological cycle \citep{Rannou06}, stratospheric gases \citep{Hourdin04}, and haze-dynamical feedbacks \citep{Rannou04,Crespin08}. Since the Cassini spacecraft's present exploration of the Saturnian system, new GCMs developed to better take advantage of the increasing quality and quantity of data---in particular by returning to being three-dimensional---, have had success in reproducing some of the observations, but have also been encumbered by a combination of numerical difficulties and unrealistic assumptions.
\par
GCMs used to investigate Titan's methane cycle in detail \citep{Mitchell06,Schneider12} have shown that the observed distribution of clouds \citep{Rodriguez09,Brown10,Rodriguez11,Turtle11a} is a natural result of Titan's changing seasons, and that the circulation efficiently transports methane poleward \citep{Mitchell06,Mitchell12,Schneider12}, drying the equatorial regions \citep{Mitchell08}. The use of gray radiative transfer in these models, though, results in unrealistic surface insolation distributions \citep{Lora11}, and precludes extension of the models to the stratosphere.  A variety of additional simplifications, such as prescribed surface-level relative humidity and infinite methane supply from the surface, have also been employed \citep{Rannou06,Tokano09,Mitchell11,Mitchell12}, limiting their ability to predict the distribution of liquids.
\par
Furthermore, with the exception of the CAM Titan model \citep{Friedson09}, which did not simulate the methane cycle but produced a realistic temperature profile, other Titan GCMs \citep{Hourdin95,Tokano99,Richardson07}, including those used to study stratospheric dynamics and haze \citep{Rannou02,Lebonnois12}, have employed versions of the radiative transfer model of \citet{McKay89}, which works well for the troposphere and lower stratosphere, but seems to produce unrealistic temperatures higher up (including a cold stratosphere, a sharp increase in temperature in the uppermost regions, and a failure to obtain a stratopause \citep{Lebonnois12}). 
\par
Separately, an important numerical problem, namely models' inability to properly attain atmospheric superrotation \citep{Tokano99,Richardson07,Friedson09}, has been widely studied \citep{Newman11,Lebonnois12b} but remains incompletely understood. Consensus on the mechanism of its maintenance and on numerical obstacles is not evident, as only some three-dimensional models simulate it, and only under special circumstances \citep{Newman11,Lebonnois12}.
 \par
In this paper, we present simulations from the Titan Atmospheric Model (TAM), a new three-dimensional Titan GCM developed to alleviate some of these difficulties and to incorporate and study processes and phenomena being unveiled by the Cassini mission in Titan's middle and lower atmosphere. A previous lower-atmosphere version of this model was used to investigate Titan's recent paleoclimate \citep{Lora14}. The model and methodology are described in Section~\ref{Sec:model}. In Sections~\ref{Sec:middle} and \ref{Sec:lower}, simulations from the middle and lower atmosphere are benchmarked against a variety of observational constraints (temperatures, winds, humidity, and cloud locations), as well as used to explore model sensitivities. In Section~\ref{Sec:discussion}, we discuss model limitations, providing groundwork for future development and studies. We summarize relationships between observed and modeled phenomena and conclude in Section~\ref{Sec:conclusions}.

\section{Model}\label{Sec:model}
\subsection{Description of the GCM}
The GCM, which makes use of the Geophysical Fluid Dynamics Laboratory's (GFDL) Flexible Modeling System (FMS) infrastructure, couples a physics package based on GFDL's atmospheric component models to the fully three-dimensional FMS spectral dynamical core \citep{Gordon82}. Here we describe the component modules of the physics package.
\subsubsection{Radiation}
The radiative transfer model is intended to compute accurate radiative heating rates without approximations that compromise their overall vertical or latitudinal distributions. Solar-wavelength ($<$4.5 $\mu$m) and thermal infrared ($>$4.5$\mu$m) fluxes are computed employing nongray, multiple scattering, plane-parallel two-stream approximations from scaled extinction optical depths, single scattering albedos, and asymmetry parameters \citep{Toon89,Briegleb92}. Seasonal and diurnal cycles are included in the computation of insolation.
\par
Methane opacities at wavelengths short of 1.6 $\mu m$ are calculated with exponential sum fits to transmissions, using DISR absorption coefficients \citep{Tomasko08a} with varying column abundance. The effects of methane opacity between 1.6 and 4.5$\mu m$ are accounted for using correlated $k$ coefficients calculated from HITRAN line intensities \citep{Rothman09}. For the purposes of radiative transfer, the methane profile is globally set to that measured by Huygens \citep{Niemann05}.
\par
Opacities due to CIA---which include combinations of N$_2$, CH$_4$ and H$_2$ pairs---are calculated from HITRAN data \citep{Richard11} with exponential sum fits to pressure- and temperature-dependent transmissions. It should be noted that the mole fraction of H$_2$ is assumed to be constant at 0.1\% \citep{Tomasko08c}. Molecular absorption is treated with correlated $k$ coefficients from temperature- and pressure-corrected \citep{Rothman96} HITRAN line intensities. Included absorbers are CH$_4$, C$_2$H$_2$, C$_2$H$_4$, C$_2$H$_6$, and HCN. The profiles of the stratospheric molecular species are fixed to observed values \citep{Vinatier07}.

Shortwave haze optical parameters are those published by the DISR team \citep{Tomasko08b}, thus assuming that the haze distribution is horizontally homogeneous. Values of optical depth for wavelengths larger than 1.6~$\mu$m (beyond those measured by DISR) are extrapolated using power law fits \citep{Tomasko08b}; the haze is assumed to become more absorptive with increasing wavelength beyond 1.6~$\mu$m. Haze in the thermal infrared is assumed to be perfectly absorbing, so that the single scattering albedo is always zero. Optical depths are calculated from volume extinction coefficients determined from Cassini/CIRS data, available between wavenumbers 20--560~cm$^{-1}$ \citep{Anderson11} and 610--1500~cm$^{-1}$ \citep{Vinatier12}. Values for wavenumbers larger than 1500~cm$^{-1}$ are interpolated between these and the DISR results, using a power law fit (note that very little energy is transmitted at these wavenumbers). These volume extinction coefficients are assumed constant between 0 and 80~km, and decreasing with a scale height of 65~km above that.

\subsubsection{Moist processes}
Methane saturation vapor pressure is calculated either over an 80/20 CH$_4$/N$_2$ liquid \citep{Thompson92} or pure methane ice \citep{Moses92} depending on temperature, with the transition at 87 K where the vapor pressure curves intersect. The effects of ethane on the vapor pressure of methane are assumed negligible.
\par
Two precipitation schemes are included: A large-scale condensation (LSC) scheme, which condenses any methane per grid box exceeding 100$\%$ relative humidity and allows it to re-evaporate in underlying layers, and a quasi-equilibrium moist convection scheme \citep{Frierson07,O'Gorman08}, where convectively unstable columns relax toward a moist pseudoadiabat. In the latter, excess liquid falls immediately to the surface. In both cases, whatever condensation occurs is assumed to be liquid, ignoring the $\sim$10$\%$ difference in latent heats between ice and liquid, avoiding the need to model the ice-liquid transition for energy balance. Furthermore, in all cases it is assumed that nucleation is always possible, ignoring detailed microphysics. The effects of clouds are neglected in the radiative transfer.
\subsubsection{Surface}
The GCM employs a soil model using 15 layers of variable thickness to 80 m depth, between which heat is transported by conduction. The thermal properties of the soil are assumed to be those appropriate for the ``porous icy regolith" of \citet{Tokano05}. Neither topography nor albedo variations are included. At the ground surface, fluxes of sensible and latent heat, radiation, and momentum are calculated using bulk aerodynamic formulae, with drag coefficients from Monin-Obukhov similarity theory. Roughness length and gustiness parameters in this module are assumed to be 0.5 cm and 0.1 m~s$^{-1}$, respectively \citep{Friedson09,Schneider12}.
\par
A ``bucket" model tracks the liquid content of the ground $q_{g}$,
\begin{equation}
\frac{\partial q_{g}}{\partial t} = P - E,
\end{equation}
where $P$ is precipitation---resulting from moist processes---that accumulates on the surface, and $E$ is evaporation that removes methane from the surface reservoir. An availability factor parameterizes infiltration and sub-grid scale ponding, limiting evaporation when the grid box has less than 100 kg m$^{-2}$ of methane and linearly decreasing to zero evaporation at zero methane. The thermal effects of liquid methane are not included, and surface liquid cannot move laterally, even when multiple grid boxes are necessary to define a ``lake" or ``sea." 
\subsubsection{Boundary layer}
Vertical diffusion in the boundary layer uses a standard K-profile scheme, wherein diffusivities of heat and moisture, $K_H$, and momentum, $K_m$, are calculated as a function of height $z$ within the boundary layer $h$, and Monin-Obukhov stability functions $\Phi_{H,m}$: 
\begin{equation}
K_{H,m} = 
	\begin{cases}
		(ku_*z/\Phi_{H,m}) & \text{for } z<h_{low}\\
		(ku_*z/\Phi_{H,m})\times\left(1-\frac{z-h_{low}}{h-h_{low}}\right)^2 & \text{for } h_{low}\leq z < h,
	\end{cases}
\end{equation}
where $k$ is the von Karman constant, $u_*$ the surface friction velocity calculated from Monin-Obukhov theory, and $h_{low}$ the height of the surface layer, assumed to be one-tenth of the boundary layer height. For stable or neutral conditions, the boundary layer height $h$ is set where the Richardson number, the ratio of potential to kinetic energy and a measure of the importance of buoyancy, equals 1.0. In the case of unstable conditions, $h$ is set at the level of neutral buoyancy for surface parcels. 
\subsection{Methodology}
A series of simulations was carried out to examine various aspects of the model atmosphere. All simulations presented here used relatively low T21 resolution (roughly 5.6$^{\circ}$ horizontal resolution) to minimize computational requirements, and were run with eighth-order hyperdiffusion to dissipate enstrophy that builds up at the model's smallest resolved scales. An additional diffusive ($\nabla^2$) ``sponge" was applied to wind fields at the top-most layer to reduce wave reflections and improve numerical stability. Most simulations used a 32-layer, hybrid-coordinate atmosphere extending from the surface to approximately 40 $\mu$bar, hereafter referred to as L32. One simulation used 50 layers extending to about 3 $\mu$bar (hereafter L50).
\par
In order to test the model's ability to naturally super-rotate, a ``control," L32 simulation was started from rest (zero wind speeds globally) and allowed to run until equilibrium was reached in the atmospheric variables. The superrotation stabilized after about 70 Titan years of integration. This simulation was run for an additional five Titan years, and all other L32 simulations were initialized with this spun-up atmosphere. A continuation of the control simulation was used for any direct comparisons. 
\par
Once the model's capacity for superrotation was established, a much more computationally expensive L50 run was started from a prescribed superrotating state (the timestep was reduced to six from 15 minutes). This simulation's superrotation stabilized quickly, within two Titan years. An additional two years were run for analysis.
\par
A L32 simulation to briefly test the effects of a variable haze versus the control simulation was run with the following parameterization: After computation of the haze optical depth in the radiative transfer module, each layer's haze optical depth $\mathrm{d}\tau$ above an altitude of 80 km was modified as
\begin{equation}
\mathrm{d}\tau = \mathrm{d}\tau \left(1-(\cos(2\phi)-1)|\sin(t)|\right),
\end{equation}
where $\phi$ and $t$ are latitude and orbital time, respectively. Thus, a seasonally oscillating enhancement of wintertime haze optical depth, more pronounced at the poles, represents a rough parameterization of haze transport by the atmosphere. 
\par
Finally, two additional simulations were used to examine the methane cycle in the lower atmosphere (without parameterized haze variability). In the first, the surface methane was replaced with a deep reservoir of 100 m, which represents an inexhaustible global surface methane reservoir akin to what has been used in most previous studies \citep{Mitchell06,Rannou06,Tokano09,Mitchell12}. This simulation reached equilibrium in less than five Titan years, and was run for an additional five. For the second, a global reservoir of four meters of methane was imposed, with the addition of 100 m deep reservoirs at the approximate locations of Titan's observed Ontario Lacus, Kraken Mare, Ligeia Mare, and Punga Mare (henceforth lakes/seas simulation). This simulation was run for considerably longer (35 Titan years), as the surface reservoir only stabilized after the mid- and low-latitude surfaces dried.

\section{Middle atmosphere}\label{Sec:middle}
In this section, we discuss results from three simulations: the L50 case extending into the mesosphere, the L32 control case, and the L32 simulation using parameterized varying haze.

\subsection{Atmospheric temperatures}
\begin{figure}[h!]
	\begin{center}
	\includegraphics[width=1.0\textwidth]{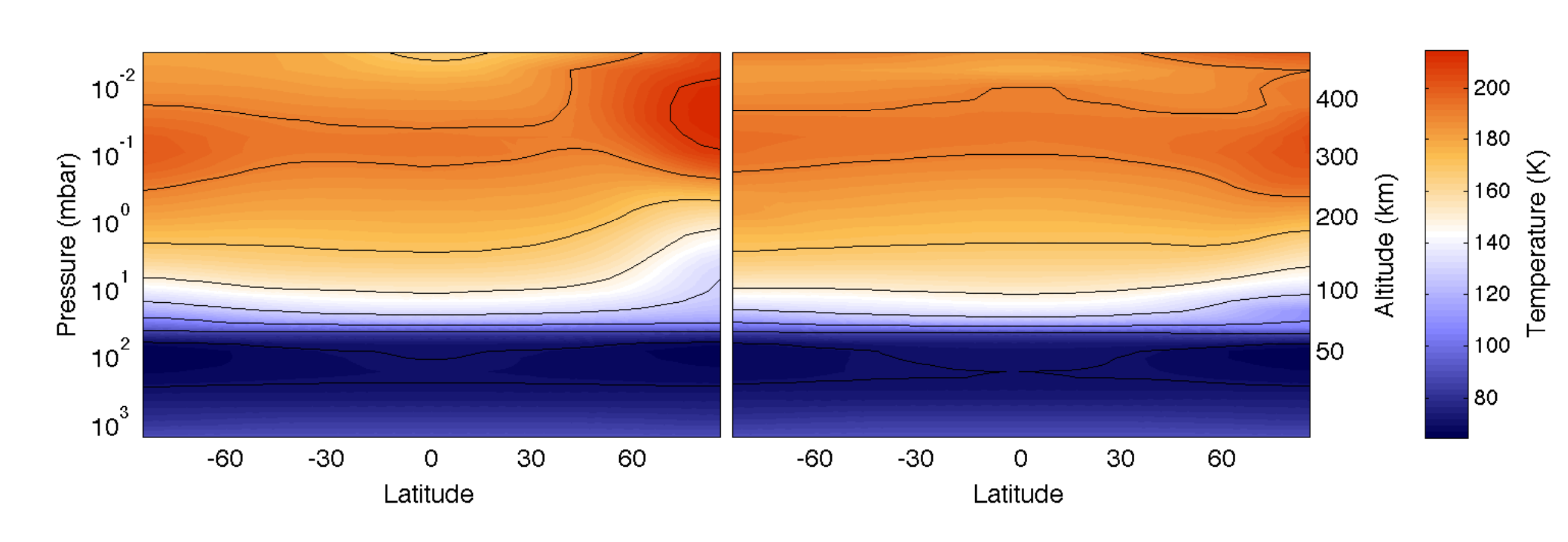}
	\caption[Zonal-mean temperatures]{Zonal-mean temperatures (K) from (left) northern fall, $L_S\sim$230$^{\circ}$, and (right) northern winter, $L_S\sim$315$^{\circ}$, from L50 simulation. An approximate altitude scale is included for reference.\label{Fig:temps_50L}}
	\end{center}
\end{figure}

Zonally averaged temperatures from the L50 simulation are shown in Fig.~\ref{Fig:temps_50L}. The times shown correspond to the seasonal extrema of the superrotation during northern fall and winter. An immediately apparent feature is the existence of a clear stratopause at all latitudes, at pressures of around 0.03 to 0.1 mbar. Though this occurs at a somewhat lower pressure than the stratopause observed by the Huygens probe \citep{Fulchignoni05}, it is in excellent agreement with Cassini CIRS observations of the middle atmosphere \citep{Flasar05,Achterberg08}, in which the stratopause occurs roughly between 0.05 and 0.1 mbar.

\par
The warmest temperatures occur directly over the winter polar regions, as a result of adiabatic heating of descending air driven by the meridional circulation. This warm region appears at pressures below those of the rest of the stratopause. As winter progresses into spring, it then descends in altitude and cools, decreasing the contrast with the low-latitude stratopause. The initial altitude and subsequent cooling agree well with thermal emission spectral data \citep{Achterberg11}. However, the timing of the simulated process is different than those observations: A warm polar region of about 210 K was seen on Titan shortly \textit{after} northern winter solstice, whereas in the model that feature is already dissipating at the corresponding time. This discrepancy may be due to the modeled haze and radiatively active stratospheric gases being uncoupled to the dynamics and horizontally homogeneous, as this warm feature is a balance between adiabatic heating and radiative cooling.
\par
\begin{figure}[h!]
	\begin{center}
	\includegraphics[width=0.8\textwidth]{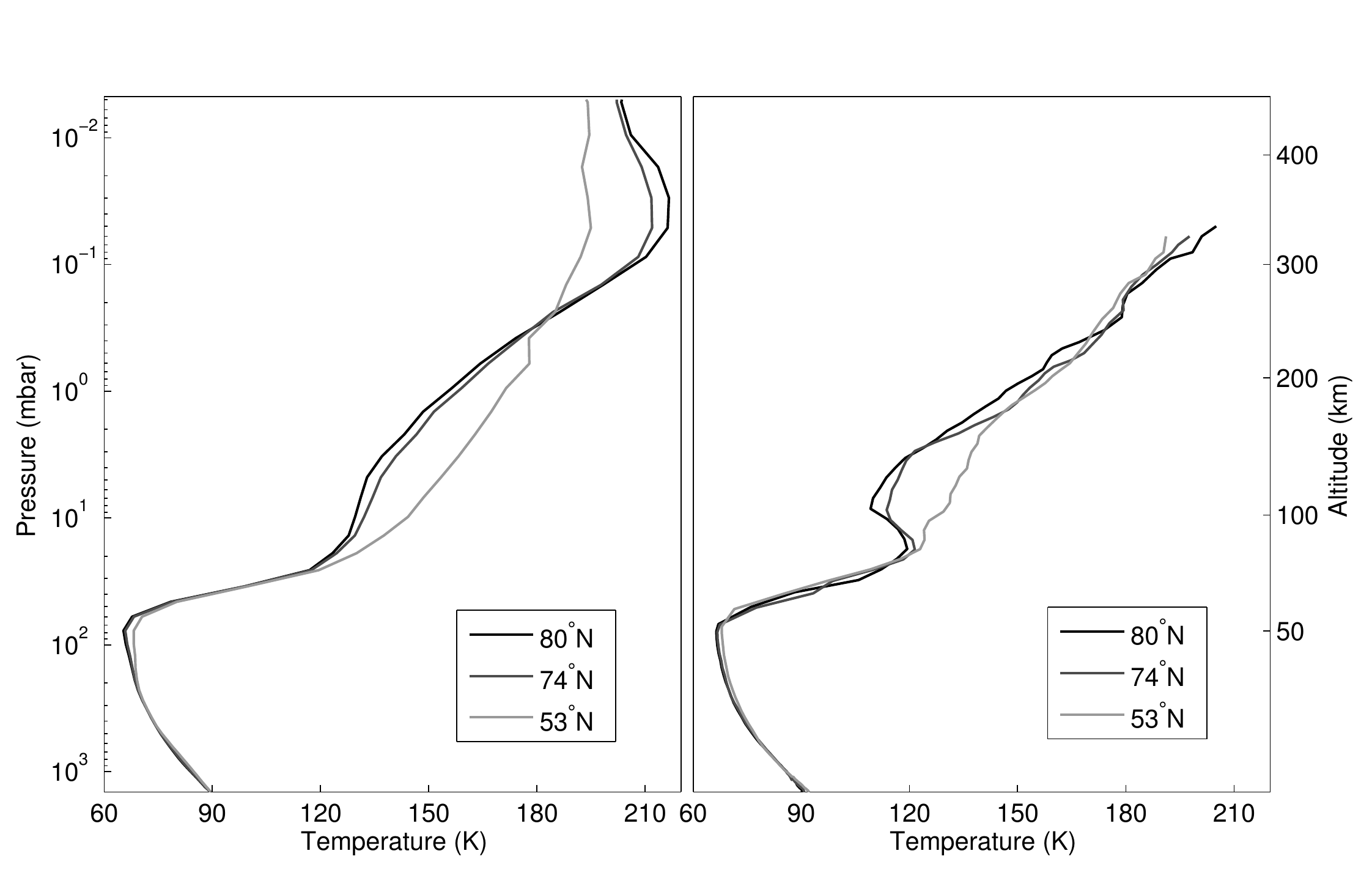}
	\caption[Polar temperature profiles]{Left: L50 simulated vertical temperature at three latitudes shortly before winter solstice. Right: Selected radio occultation temperature profiles from the highest observed latitudes, during various times in late northern winter \citep{Schinder12}. An approximate altitude scale is included for reference.\label{Fig:polar_temps}}
	\end{center}
\end{figure}

In the lower stratosphere, high-latitude winter regions are coldest, with low- and summer latitudes having relatively flat isotherms, also agreeing with thermal emission spectral data \citep{Achterberg08}. The cold winter polar regions in this part of the atmosphere may be in part the result of radiative cooling during the polar night (Titan's effective obliquity is 26.7$^{\circ}$). A rather drastic cooling of the atmosphere around 1--10 mbar seen particularly in the left panel of Fig.~\ref{Fig:temps_50L} occurs as the stratopause above it warms. This is shown as vertical profiles in Fig.~\ref{Fig:polar_temps}. At around the same time that adiabatic heating begins to warm the polar stratopause, a significant cooling of the stratosphere around 0.3 mbar occurs, coinciding almost exactly with the onset of polar night. The cooling slowly propagates downward and equatorward. The resulting temperature oscillation slowly extends to higher pressures over the course of the season, reaching approximately the 5 mbar level around $L_S\sim$230$^{\circ}$. This feature is consistent with high-latitude temperature profiles observed from radio occultations \citep{Schinder12}, which display a temperature inversion in the middle stratosphere. Though the simulation does not develop a proper inversion, the upper stratospheric temperatures, the sharp change in temperature gradient, the pressure where the feature again joins the ``background" temperature profile, and the variation with latitude are all remarkably similar. This feature in the model also seems to dissipate too early, and has practically disappeared by the time corresponding to the actual observations. A likely candidate for this too-fast warming may be the lack of buildup of stratospheric gases in the winter vortex. The altitude drop of the stratopause in late winter, not apparent in observations, is probably closely related. Note also that in the L32 simulation, which does not develop a proper stratopause, this temperature oscillation is almost non-existent, indicating a strong connection between the two features, probably related to their radiative effects.

\par
In the troposphere, temperatures at all latitudes are in excellent quantitative agreement with observations. At mid and low latitudes (between $\pm60^{\circ}$), the tropopause remains between 69.5 and 71 K year-round, while the high-latitude tropopause dips to $\sim$64 K in winter and also occurs at lower pressures, as seen by radio occultations \citep{Schinder12}. This tropopause temperature is also a sharper minimum than at lower latitudes where the tropopause is bracketed by a region that is nearly isothermal. Both of these features agree with observations.
\par
\begin{figure}[h!]
	\begin{center}
	\includegraphics[width=0.5\textwidth]{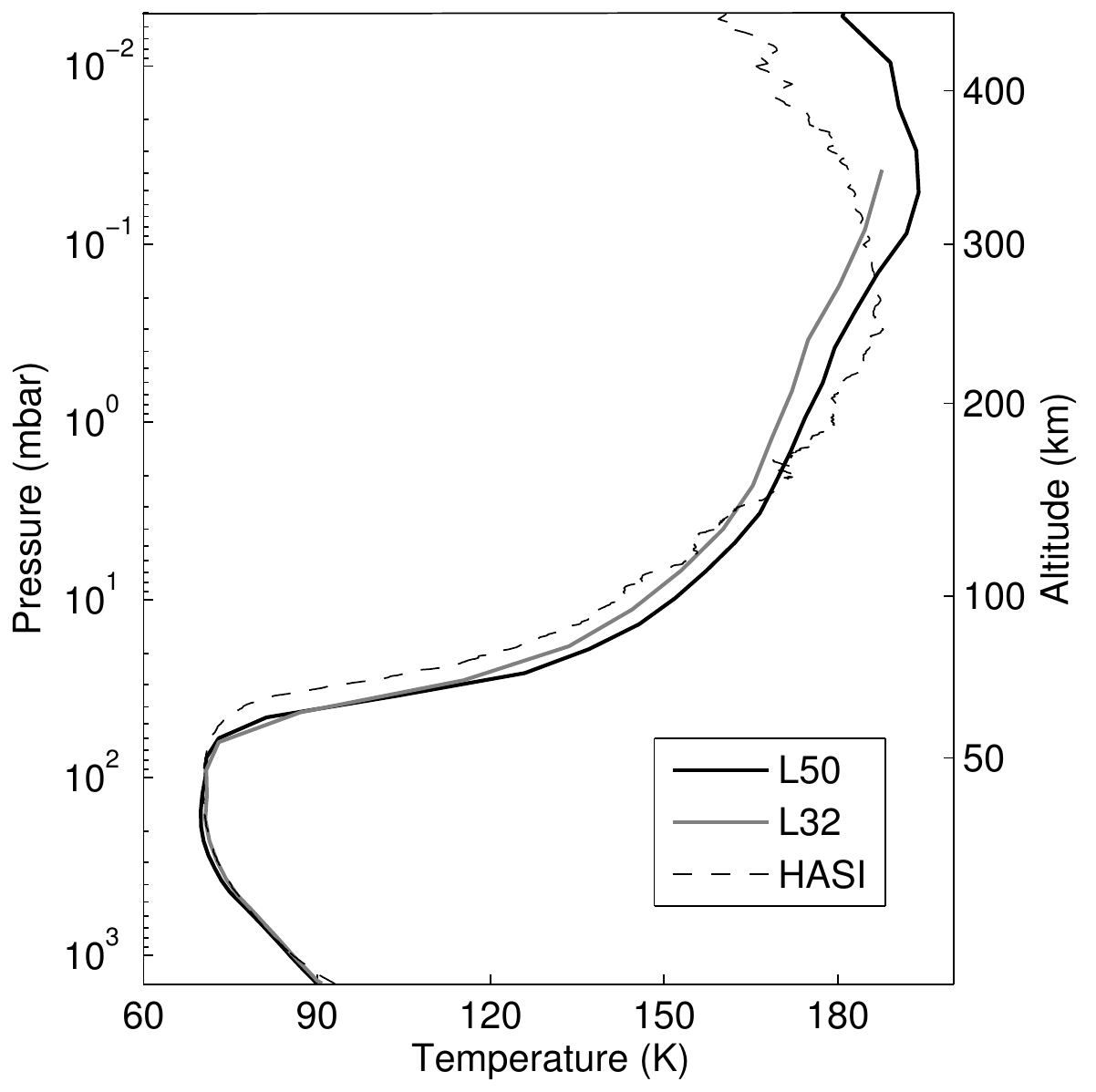}
	\caption[Equatorial temperature profile]{Simulated and observed \citep{Fulchignoni05} vertical temperature profile near the equator for the time of the Huygens descent. An approximate altitude scale is included for reference. \label{Fig:temp_profiles}}
	\end{center}
\end{figure}

A comparison of the vertical temperature structure from both L50 and L32 control simulations, relevant to the time and season of the Huygens probe's descent, is shown in Fig.~\ref{Fig:temp_profiles}. In both cases, the lowest levels of the stratosphere occur at too-high pressures, though in general the stratospheric temperatures are close to those observed, without qualitatively different structures appearing in the simulations. Though the Huygens observations display a stratopause at around 0.3 mbar \citep{Fulchignoni05}, the same is not true of Cassini CIRS data \citep{Flasar05} or radio occultations \citep{Schinder11}; our results are in excellent agreement with these observations, as discussed above, all the way through the model domain. Note that no tuning of radiative parameters was done to achieve these temperature profiles.

\par
Between the low stratosphere and the surface, the agreement between the simulations is good. Toward the high stratosphere, the L32 simulation is increasingly cold in comparison, probably due to the model top inhibiting the formation of a full stratopause. Nevertheless, the two simulations are in adequate agreement, validating the use of the L32 model for investigating the lower stratosphere and troposphere. 

\subsection{Meridional Circulation}
\begin{figure}[h!]
	\begin{center}
	\includegraphics[width=0.8\textwidth]{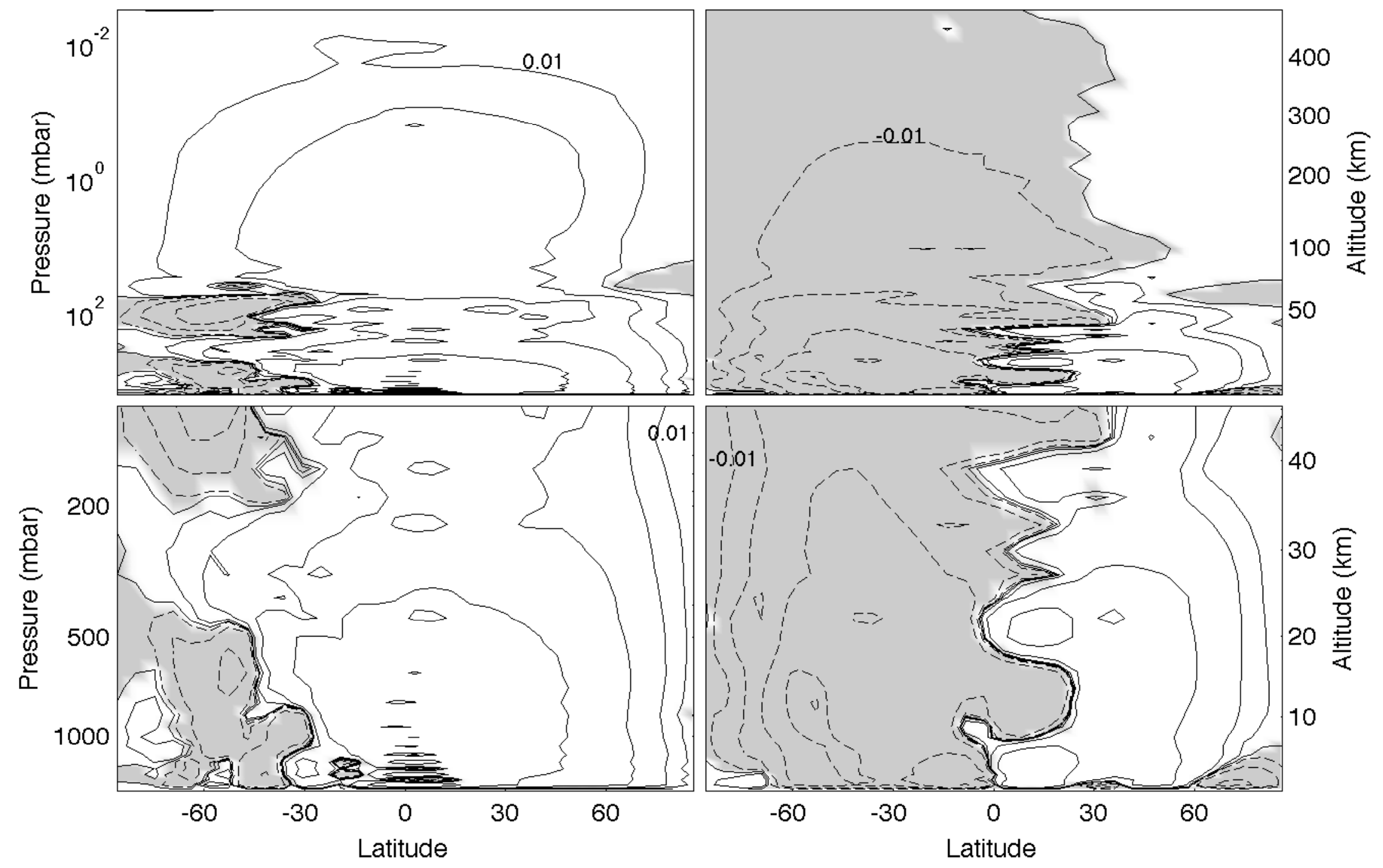}
	\caption[Streamfunctions]{Mean meridional streamfunction (10$^9$ kg~s$^{-1}$) corresponding to southern summer solstice (left) and northern spring equinox (right), showing the mean circulation from the L50 simulation. Bottom panels show a zoomed view of the troposphere. Positive values indicate clockwise motion; negative values are dashed lines in shaded regions. The contour magnitudes increase by factors of 4, with the lowest magnitude labeled. An approximate altitude scale is included for reference.\label{Fig:streamfunction}}
	\end{center}
\end{figure}

The zonal-mean meridional streamfunction of the L50 simulation is shown in Fig.~\ref{Fig:streamfunction} for southern summer solstice ($L_S\sim$270$^{\circ}$) and northern vernal equinox ($L_S\sim$0$^{\circ}$). In the former case, a pole-to-pole Hadley circulation is apparent, particularly in the stratosphere, with rising motion in the summer hemisphere and subsidence in the winter hemisphere. This is consistent with previous models \citep[e.g.,][]{Friedson09,Newman11,Lebonnois12}. A small tropospheric cell is also visible at high southern latitudes, due to rising motion occurring over the location of maximum surface heating; this is further discussed in the following section. Previous three-dimensional models show similar structures at high latitudes \citep{Newman11,Lebonnois12}, and \citet{Mitchell09} also showed in an axisymmetric model that latent heating limits the Hadley upwelling in the troposphere, similarly to our simulated circulation. During equinox, a more symmetric equator-to-pole circulation develops throughout the atmosphere, as the equivalent of an intertropical convergence zone (ITCZ), where rising motion dominates, crosses the equator. 
\par
This meridional circulation (Fig.~\ref{Fig:streamfunction}) is also representative of that from the L32 simulations, though in those cases the lower model top expectedly suppresses the circulation of the lowest pressure levels.

\subsection{Zonal Winds}
A primary aim of this model was to both reproduce the temperature structure through the stratopause and also achieve atmospheric superrotation, something that has proven difficult for three-dimensional models \citep[e.g.][]{Friedson09,Newman11,Lebonnois12,Tokano13}. In the L32 control simulation spun up from rest, the atmosphere quickly becomes superrotating, though with zonal wind magnitudes lower than the observed $\sim$200 m~s$^{-1}$ \citep{Achterberg08}, of around 130 m~s$^{-1}$ \citep[similar to what was attained by][]{Hourdin95}. Strong winter jets develop within the first few Titan years of simulation, with maximum wind speeds over high mid-latitudes. During spring/fall, the wind maximum travels across the equator, dissolving the springtime polar jet and ramping up the opposite hemisphere's. During this time, the peak winds also shift to lower pressures, in agreement with observations, which suggest an increase in windspeeds of several tens of m~s$^{-1}$ at pressures below 0.1~mbar between 2005 and 2009 \citep{Achterberg11}. The maximum integrated angular momentum occurs shortly after equinoxes, when wind speeds also peak. 
\par
All latitudes in the stratosphere and upper troposphere continuously support westerlies. In the middle troposphere, zonal winds reach tens of meters per second, in agreement with inferred winds from cloud observations \citep{Griffith05,Porco05}. Close to the surface, easterlies dominate at low latitudes, with mid-latitude winds oscillating between pro- and retro-grade with season.
\par
\begin{figure}[h!]
	\begin{center}
	\includegraphics[width=1.0\textwidth]{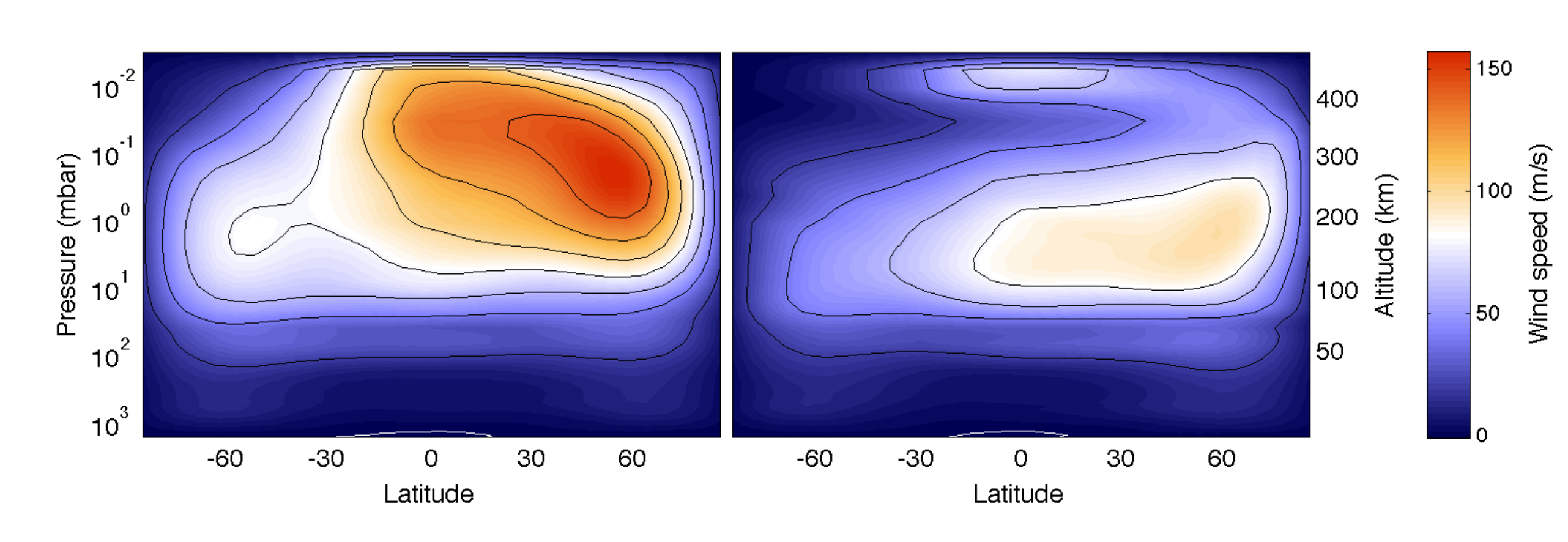}
	\caption[Zonal-mean zonal winds]{Zonal-mean zonal winds (m~s$^{-1}$) from northern mid-fall (left) and mid-winter (right) from the L50 simulation, corresponding to the temperature fields in Fig.~\ref{Fig:temps_50L}. The contour interval is 20 m~s$^{-1}$; the 0 m~s$^{-1}$ contour is in white. An approximate altitude scale is included for reference. \label{Fig:winds_50L}}
	\end{center}
\end{figure}

The behavior of winds in the L50 simulation is similar, though with slightly higher (more realistic) wind speeds. Superrotation extrema lag in comparison to the L32 by about 15$^{\circ}$ of $L_S$, with peak winds occurring during mid-fall (Fig.~\ref{Fig:winds_50L}), and weakening through winter before the hemispheric reversal. Some effects of the top-layer sponge are apparent at the model top, and the highest few layers cannot be considered reliable.    Nevertheless, the simulated winds at pressures above 0.01 mbar are satisfactory.

\par
\begin{figure}[h!]
	\begin{center}
	\includegraphics[width=0.5\textwidth]{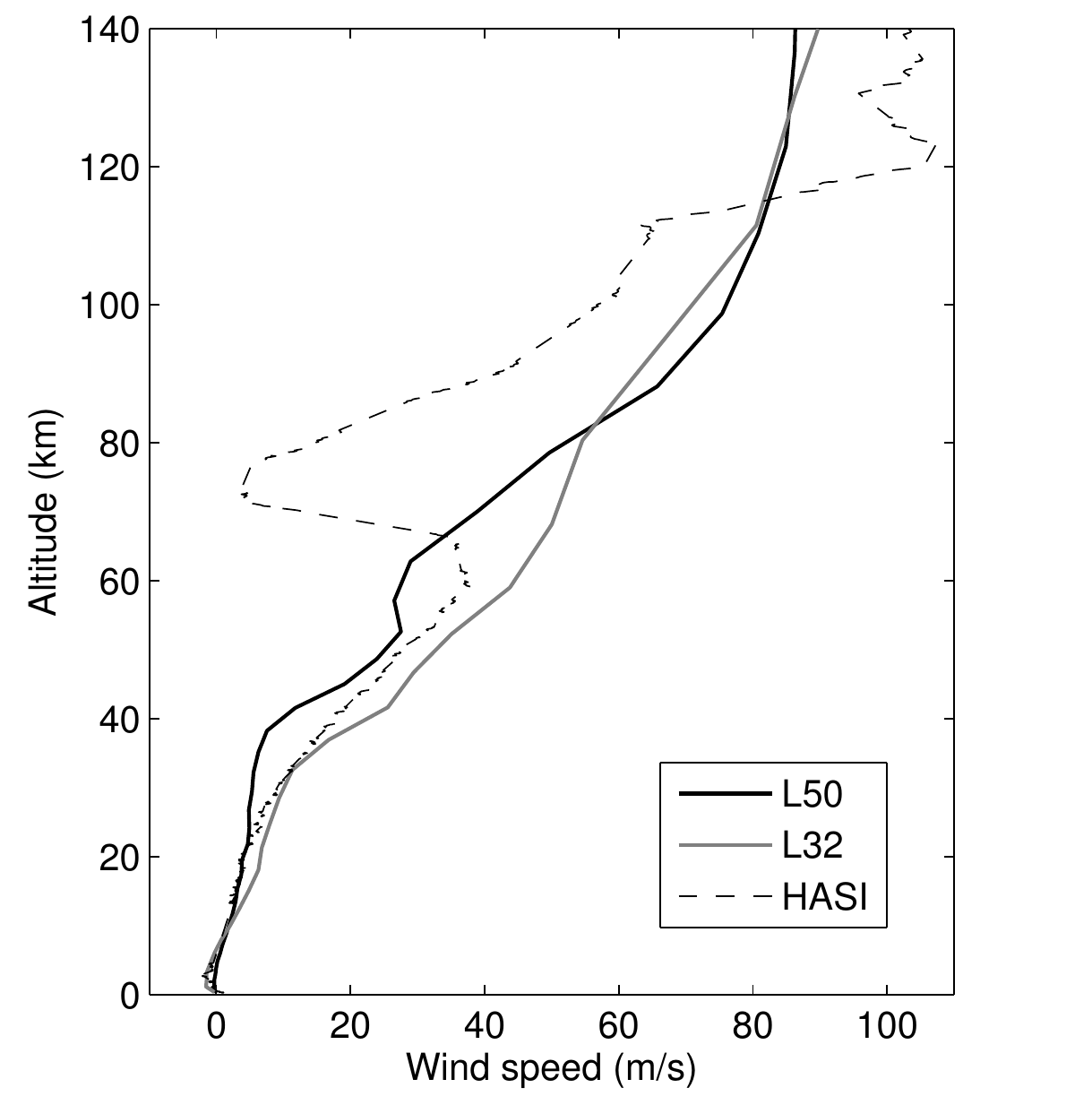}
	\caption[Equatorial wind profile]{Simulated zonal-mean zonal wind and observed zonal wind \citep{Bird05} profile near the equator for the time of the Huygens descent. \label{Fig:wind_profiles}}
	\end{center}
\end{figure}
\begin{figure}[h!]
	\begin{center}
	\includegraphics[width=0.7\textwidth]{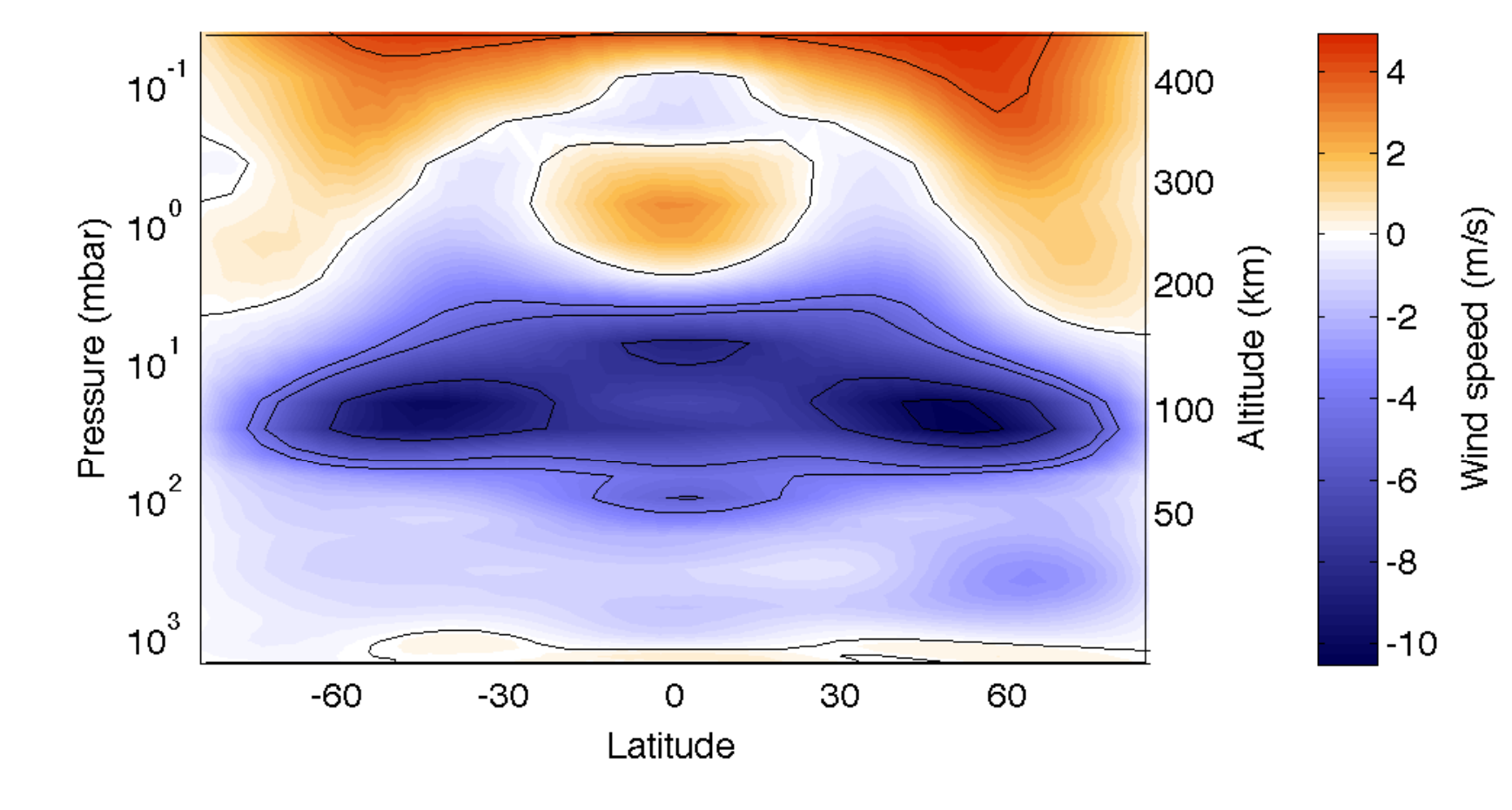}
	\caption[Variable haze, control zonal wind difference]{Difference between the L32 annual-mean variable haze and control simulations' zonal-mean zonal winds (m~s$^{-1}$). An approximate altitude scale is included for reference.
	 \label{Fig:haze-control_winds}}
	\end{center}
\end{figure}

Figure~\ref{Fig:wind_profiles} shows the simulated vertical profile of zonal wind at the season and approximate latitude of the Huygen's descent, compared to observations. The agreement in the troposphere in the control case is good, including the presence of weak easterlies between the surface and $\sim$5 km altitude (though the observed weak surface westerlies are not present \citep{Bird05}). Just above the tropopause, there is a decrease in the windspeed gradient with altitude, especially pronounced in the L50 model, but neither simulation reproduces the observed stillness between 70 and 80 km. \citet{Lebonnois12} produce a modest decrease in zonal winds in the vicinity of this region, suggesting its formation may be related to haze feedbacks (see below and Fig.~\ref{Fig:haze-control_winds}). It is also worth noting that the observed altitude of this drop-off coincides with that where the previously described polar temperature oscillation ends.

\par
Above 80 km, the wind profiles again agree satisfactorily up to the altitude of the \textit{in situ} measurements \citep{Bird05}. As stated above, the winds drop off too quickly in the upper stratosphere above this, particularly in the L32 simulation, so the peak winds observed by CIRS at $\sim$0.1~mbar \citep{Achterberg08} are not attained. This difficulty is likely due to a variety of model constraints, including the low model top and the lack of haze or trace gas coupling, but should not significantly affect the results in the troposphere and lower stratosphere.
\par

\par
The annual-mean difference between zonal-mean zonal winds from the variable haze and control simulations (both L32) is shown in Fig.~\ref{Fig:haze-control_winds}. Two main features are immediately apparent: First, in the high stratosphere, the varying haze acts to increase the wind speeds, and more so at higher latitudes (with peak instantaneous differences $>$15 m~s$^{-1}$ in the winter jets). Second, a decrease in wind speed occurs around 20 mbar. This coincides with the altitude of the observed wind speed minimum in the stratosphere. Though it is also exactly the altitude (80 km) of the chosen transition between varying and non-varying haze, tests with a lower transition (40 km; not shown) produced no difference in the altitude or magnitude of this wind deceleration, and high-altitude winds were additionally enhanced. This simple parameterization of the seasonal variability of haze is insufficient to accurately study this phenomenon, but it appears plausible that this variability may at least partially affect the apparent de-coupling of tropospheric and stratospheric winds, seen in the data \citep{Bird05}. These features illustrate the importance of the stratospheric haze on the zonal winds, which agrees with the conclusion from axisymmetric models that haze-dynamics coupling enhances, rather than suppresses, wind speeds in the stratosphere \citep{Rannou04}. Further studies of the relationship between varying haze, polar temperatures, and the paucity of winds in this region of the atmosphere will be the subject of a future study. It should be noted, however, that the difference in wind speeds between variable and non-variable haze simulations is modest, and the impact of other effects, such as resolution, also needs to be explored.

\subsection{Superrotation}
Despite the lower-than-observed wind speeds in the high stratosphere (particularly in the L32 simulations where vertical resolution is low and the circulation is affected by the top-most layer sponge), the agreement between simulated and observed zonal winds is good at least below approximately 1.0 mbar in both simulations (all of Fig.~\ref{Fig:wind_profiles}), and the simulated atmosphere is clearly adequately superrotating. This is a result of relative angular momentum build-up, which derives from surface torques that transfer net angular momentum from the solid body to the atmosphere.

\begin{figure}[h!]
	\begin{center}
	\includegraphics[width=0.5\textwidth]{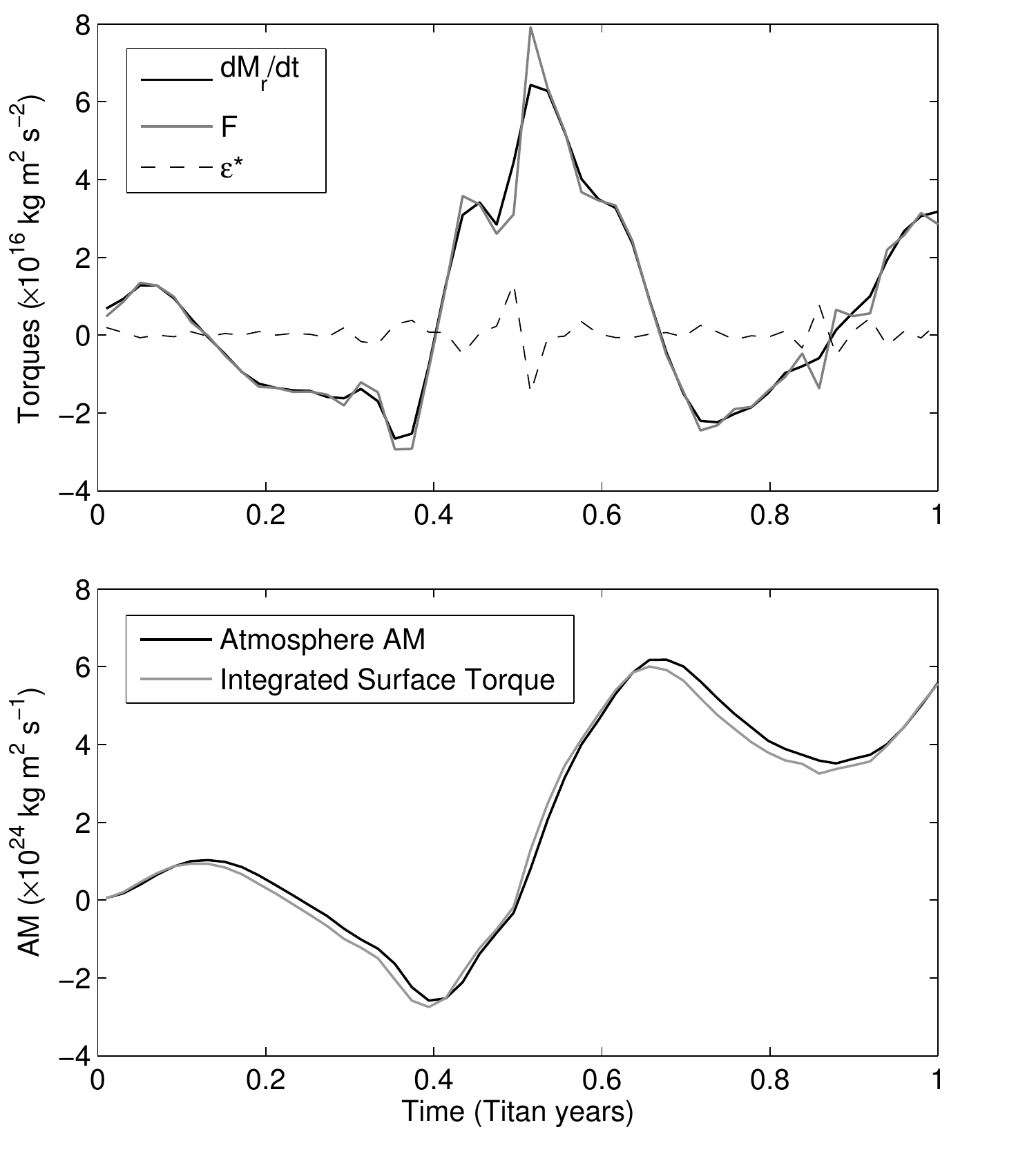}
	\caption[Torques and angular momentum budget]{Top: The rate of change of the atmospheric relative angular momentum $dM_r/dt$, net friction torque from the surface $F$, and spurious numerical torques, $\epsilon *$, for the first year of the L32 control simulation. Bottom: The corresponding total atmospheric relative angular momentum (AM) and integrated surface torque, $\int F \mathrm{d}t$. Ideally, these two curves would be identical.
	\label{Fig:AAM}}
	\end{center}
\end{figure}

\par
Figure~\ref{Fig:AAM} shows the surface torque and rate of change of the atmospheric angular momentum (top), and the total atmospheric angular momentum versus integrated surface torque (bottom), for the first year of the L32 control simulation, started from rest. The top panel also shows the total numerical torque, $\epsilon *$, which represents spurious torques due to conservation errors from the dynamical core and hyperdiffusion, as well as the effect of the top layer sponge \citep[see][]{Lebonnois12b}. Though this numerical torque is not zero (the ideal case), it remains for the most part significantly smaller than the net friction torque from the surface, and therefore does not impede the development of the atmosphere's angular momentum: the curves in the bottom panel are very close, and are positive after a year of simulation. This is further validation that the physical and numerical representation of Titan's atmosphere in our model is robust. Note that, since these simulations do not include topography, mountain torques are not simulated, though they may have an additional impact on the angular momentum budget.
\par
In our development of this Titan GCM, we initially coupled the physics package to the finite volume, cubed-sphere dynamical core from the GFDL Atmosphere Model 3 \citep[AM3;][]{Donner11}. However, we found that, with that dynamical core, the numerical torques $\epsilon *$ compensate the net frictional torques $F$ almost exactly---similarly to what is shown in Fig.~3b/d of \citet{Lebonnois12b} for a simplified-physics Venus GCM with the CAM5 dynamical core---and thereby completely prevent the buildup of atmospheric angular momentum. In our case, using a ``full" as opposed to simplified physics package, tests with basic topography, as well as various amounts of divergence damping, did not improve the situation, though these were by no means exhaustive. It is possible that similar difficulties with the CAM dynamical core, which is closely related to the GFDL core, used by \citet{Friedson09} are responsible for their failure to achieve any superrotation. Though further tests with these dynamical cores and Titan-like physics are clearly warranted, we opted to switch to GFDL's spectral core since our primary aim was a capable and realistic Titan model.

\section{Lower atmosphere and methane cycle}\label{Sec:lower}
In this section, we present results of the lower atmosphere from two L32 simulations, one with an inexhaustible, global surface liquid reservoir and another initiated with a limited surface methane supply plus deeper reservoirs at the locations of Titan's largest lakes/seas.

\subsection{Surface energy budget}
\begin{figure}[h!]
	\begin{center}
	\includegraphics[width=0.7\textwidth]{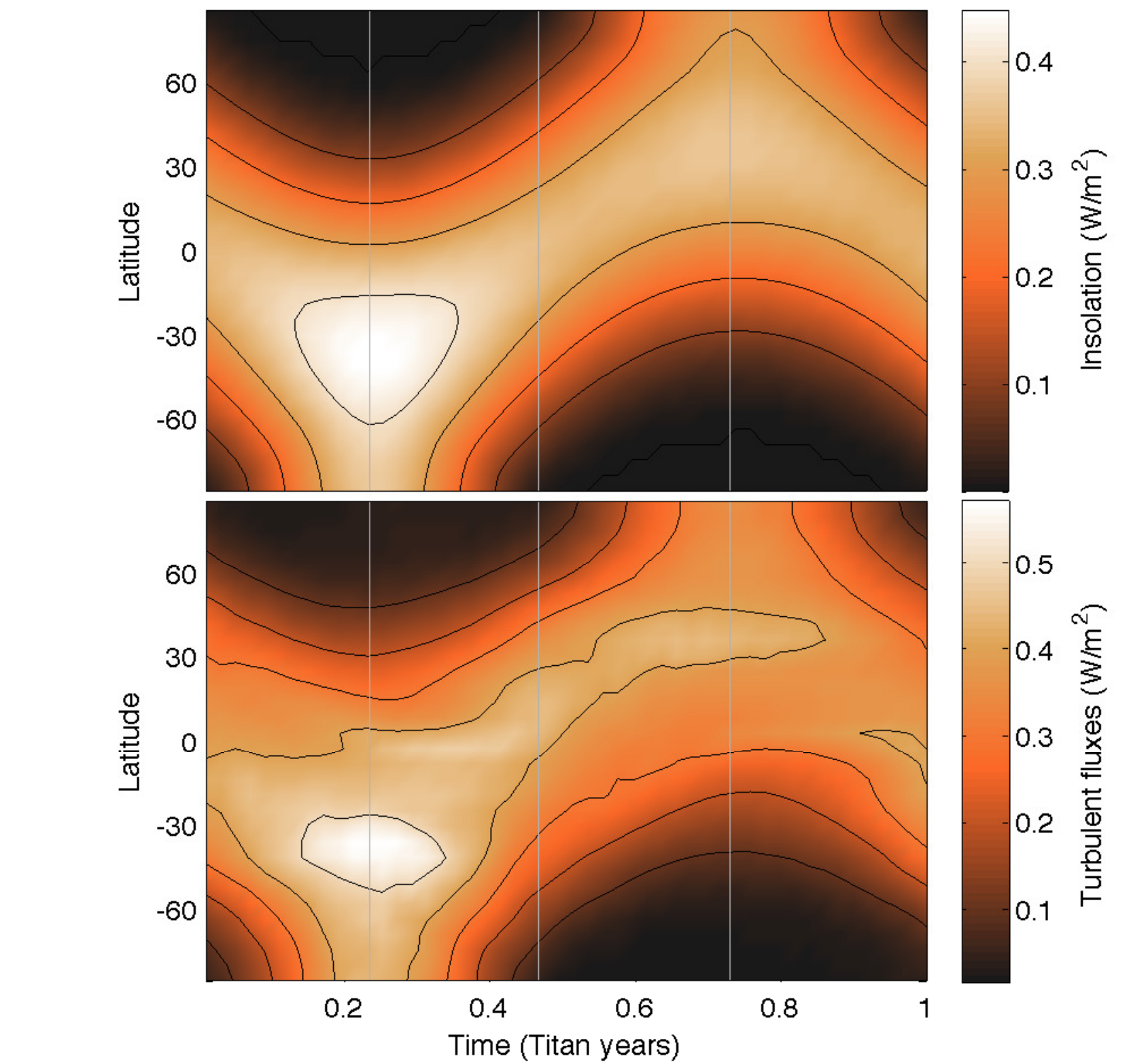}
	\caption[Surface insolation and turbulent fluxes]{Top: Insolation distribution at the surface (W m$^{-2}$). Bottom: Distribution of turbulent fluxes (sum of evaporative and sensible heat fluxes; W m$^{-2}$) from L32 global reservoir simulation. Note that the insolation is positive into the surface, while turbulent fluxes are positive into the atmosphere. Vertical lines indicate the timing of solstices and northern vernal equinox.
	\label{Fig:surf_energy}}
	\end{center}
\end{figure}

An important consequence of accurate shortwave radiative transfer is that, because of the increased pathlength through the atmosphere at high latitudes due to curvature, the distribution of insolation at the surface is not proportional to that at the top of the atmosphere \citep{Lora11}. The top panel of Fig.~\ref{Fig:surf_energy} shows this surface distribution from the GCM. Insolation peaks at summer mid-latitudes, and the southern summer, which is slightly shorter than its northern counterpart, experiences higher insolation, due to Saturn's orbital eccentricity and obliquity.

\par
The bottom panel of Fig.~\ref{Fig:surf_energy} shows the surface net turbulent fluxes, the sum of sensible heat flux to the atmosphere and evaporative (latent) energy flux, from the global surface liquid simulation. Results from the lakes/seas simulation are similar, though the partitioning between evaporation and sensible heat is entirely different. Where there is available surface methane, evaporation tends to dominate the surface flux. The distribution of these turbulent fluxes is less neatly organized than that of the insolation, but the overall pattern is still obvious, and clearly mimics the latter, with maxima at the summer mid-latitudes and minima over the winter poles. The magnitudes of these fluxes are also remarkably similar, despite the turbulent fluxes responding to the total surface radiative imbalance. (Thermal infrared fluxes, which dominate the radiative flux at the surface, are much less variable than the shortwave.) The maximum heating of the surface, often cited as the mechanism for cloud formation \citep[e.g.,][]{Brown02}, does not occur over the polar regions, and therefore neither does the maximum of destabilizing turbulent flux. Polar surface temperatures also never exceed those of the lower latitudes. Surface temperatures are further discussed below.

\subsection{Surface temperatures}
\begin{figure}[h!]
	\begin{center}
	\includegraphics[width=0.6\textwidth]{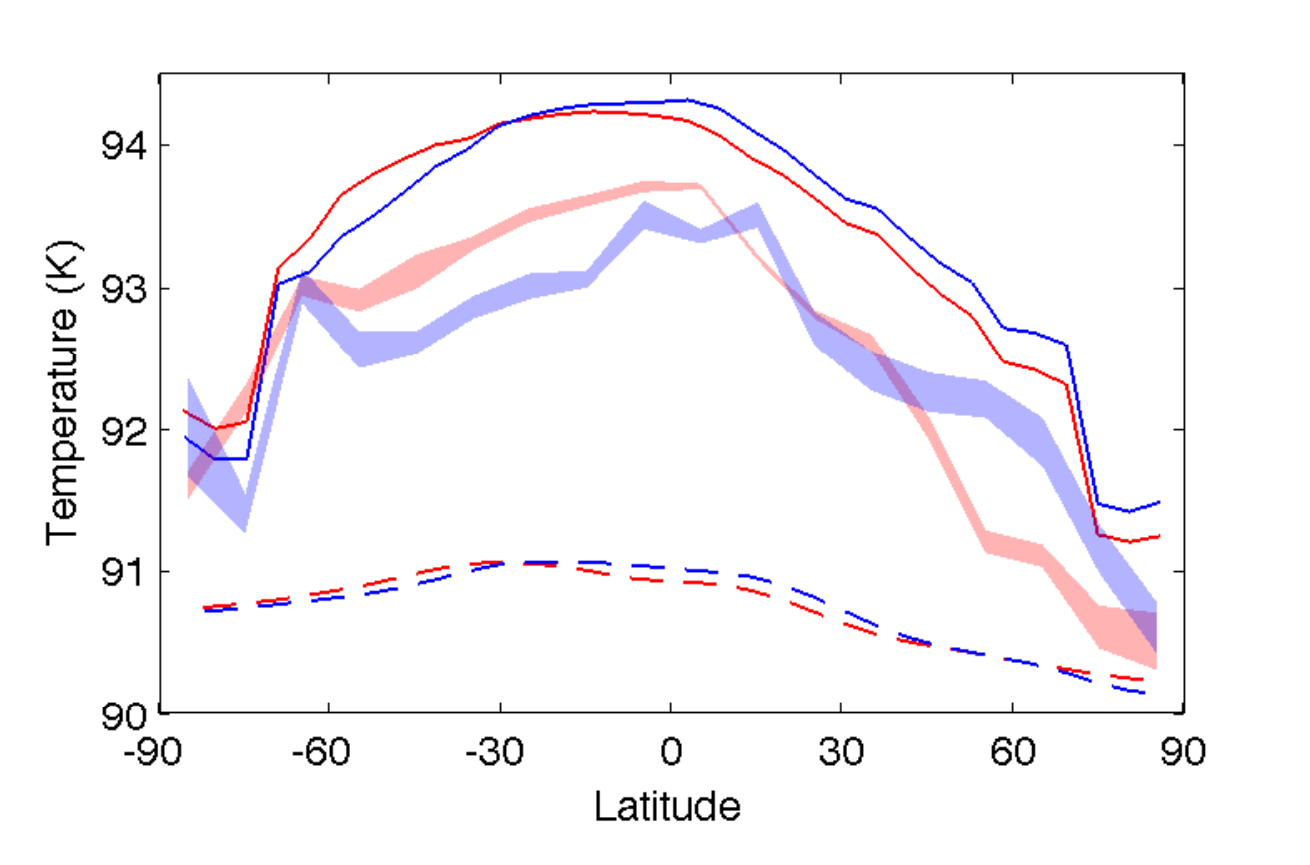}
	\caption[Surface temperatures]{Surface temperatures from two sets of L32 simulations for two seasonal periods corresponding to the thermal emission analysis of \citet{Jennings11}. Red and blue curves correspond to late northern winter (``LNW;" Sept. 2006--May 2008) and northern spring equinox (``NSE;" Nov. 2008--May 2010) periods, respectively. Solid lines are the lakes/seas simulation, dashed lines the global methane simulation, and the light shaded regions are approximately equivalent to the measurements with error bars of \citet{Jennings11}.
	\label{Fig:surf_temps}}
	\end{center}
\end{figure}

Thermal infrared measurements of Titan's surface brightness temperatures \citep{Jennings11} are compared to two sets of simulated surface temperatures in Fig.~\ref{Fig:surf_temps}. The temperatures from the lakes/seas simulation (solid lines) agree reasonably well with the observations, especially at higher latitudes. The significant decrease in temperatures poleward of $\pm70^{\circ}$, due to the prevalence of surface liquids and associated evaporative cooling, may also be present in the measurements, especially in the south, and produces a strong resemblance between simulations and measurements.
\par
Equatorward of these latitudes, the simulated temperatures are higher than observed by 0.5--1.0 K, though the observations roughly represent a zonal average that includes varying topography and different albedos and surface properties, none of which is currently included in the model. Nevertheless, the simulated surface temperatures follow the same overall trend and peak at the same approximate latitudes, roughly 10$^{\circ}$S and 5$^{\circ}$N for the two periods shown, respectively, with an equator-pole difference of about 2 K in the south and 3 K in the north. Additionally, the observed and modeled northward warming trends are in general agreement, even for a period roughly equivalent to only 10$\%$ of a Titan year. 
\par
These simulated temperatures are highest during this period, roughly during late northern winter and vernal equinox. 180$^{\circ}$ of $L_S$ later, they are generally lower (not shown), because of reduced insolation due to the larger Sun-Saturn distance. Equator-pole surface temperature gradients are approximately the same year-round, with the winter pole being coldest.
\par
On the other hand, the surface temperatures produced by the global surface liquid simulation (dashed lines in Fig.~\ref{Fig:surf_temps}) are too latitudinally homogeneous, and significantly lower than the measurements. Equator-pole contrasts in this case are only $\sim$0.2 K and 0.9 K in the south and north, respectively. These surface temperatures are the result of global evaporative cooling, and immediately highlight the implausibility of realistic simulations assuming global surface methane coverage. Note, however, that we have not varied the surface thermal properties between these two simulations. Indeed, a global ``ocean" of methane would in reality have a larger thermal inertia, so the surface temperatures would probably be even \textit{less} variable.

\subsection{Surface winds}
\begin{figure}[h!]
	\begin{center}
	\includegraphics[width=1.0\textwidth]{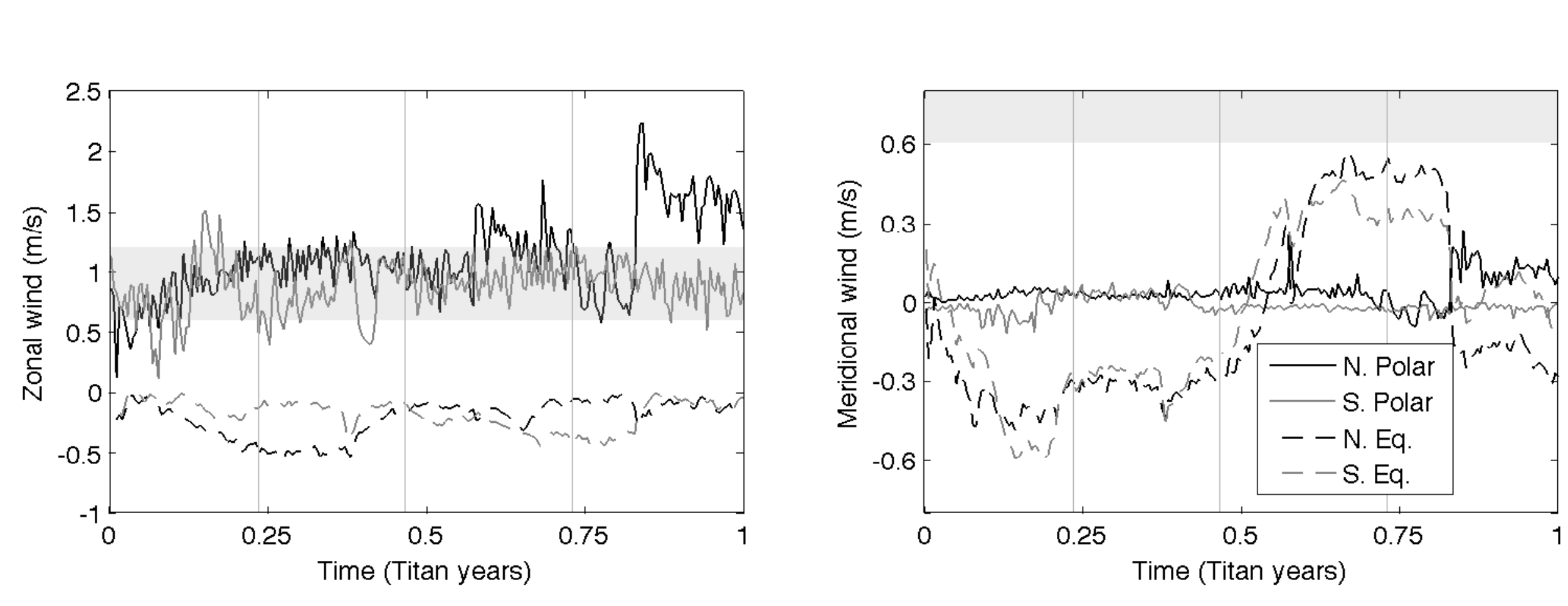}
	\caption[Surface winds]{Zonal (left panel) and meridional (right panel) winds from the lowest atmospheric layer of the L32 lakes/seas simulation. Curves labeled polar correspond to the average between 60$^{\circ}$ and 90$^{\circ}$ in each hemisphere, whereas equatorial winds (``Eq.") refer to average winds of the two gridpoints closest to the equator, roughly between 0$^{\circ}$ and 10$^{\circ}$. The legend for all curves is in the right panel. Note the different scales. The shaded areas correspond to the range of threshold wind speeds for generating waves from \citet{Hayes13}, scaled to the altitude of the simulated winds. Vertical lines indicate the timing of solstices and northern vernal equinox.
	\label{Fig:surf_winds}}
	\end{center}
\end{figure}

Figure~\ref{Fig:surf_winds} shows daily zonal and meridional wind speeds from four latitudinal regions of the GCM's lowest atmospheric layer (1439~mbar) for a Titan year of simulation with lakes/seas. Polar zonal winds are consistently eastward and stronger than equatorial winds, which are predominantly westward. In summertime, the former occasionally reach speeds approximately twice as fast as the wintertime average, which is also slightly higher in the north than the south. Though the maximum speeds increase during spring, there is considerable variability throughout the year. Meridional winds in the polar regions are at least an order of magnitude weaker, and vary significantly more during their respective hemisphere's summer, reaching their maximum magnitudes.
\par
Equatorial winds experience less variability, and the zonal component displays the opposite trend in magnitude as at the poles: Faster (easterly) winds occur during wintertime in both hemispheres, with summertime wind speeds averaging close to zero m~s$^{-1}$. On the other hand, meridional equatorial winds are of equivalent magnitude as zonal winds, and oscillate between southward and northward through a Titan year, with transitional periods of close to zero wind speed near equinoxes. Also evident is the fact that these meridional winds are cross-equatorial (flowing from winter to summer hemispheres), as the two curves vary together.
\par
We briefly discuss the implications of these wind results. In general, the seasonally reversing equatorial meridional winds are in agreement with the results of \citet{Tokano10}, and therefore with that assessment of dune orientation. Though evidence of the fast equinoctal westerlies discussed in that paper is absent here, we did not analyze instantaneous maximum and minimum wind speeds, and therefore their signal may be lost to time averaging. The variability of surface winds also suggests a connection to weather events, which could be the source of eastward gusts that may control dune orientation \citep{Lucas14}. Regardless, persistent equatorial westerlies, previously discussed as a candidate from the dune orientations \citep[e.g.][]{Radebaugh08}, are conclusively inconsistent with our results (and indeed prevalent easterlies are necessary for the flux of angular momentum into the atmosphere from the surface).
\par
Separately, our simulated polar winds are considerably stronger and somewhat more variable than those cited in Fig. 5 of \citet{Hayes13}, and the transition from below to above the threshold speed for generating waves on Titan's seas is not evident. Our results imply yearly mean reference height (10 m) winds of $\sim$0.5 m~s$^{-1}$, just around the threshold speeds suggested. However, especially in the northern hemisphere, occasional winds exceeding 2 m~s$^{-1}$ (equivalent to reference height winds $\sim$1.3 m~s$^{-1}$) occur in summertime, and are well above the thresholds; thus, our results agree with the prediction of waves on Titan's northern seas in summertime. Both zonal wind speeds and the slight increase after northern vernal equinox (roughly 0.55 of one Titan year) are in agreement with the possible detection of waves on Punga Mare, and the inferred wind speed of $\sim$0.8 m~s$^{-1}$ \citep{Barnes14}. Nevertheless, the onset of wave activity is not obvious from the simulations, and the detection of waves may depend on the timing of observations, as the wind speeds are not persistently high during the season. It is furthermore unclear that the model's low resolution is capable of predicting the appropriate mesoscale conditions that might be the dominant influence on local wave-generating wind speeds.

\subsection{Humidity and methane cycle}
\begin{figure}[h!]
	\begin{center}
	\includegraphics[width=0.5\textwidth]{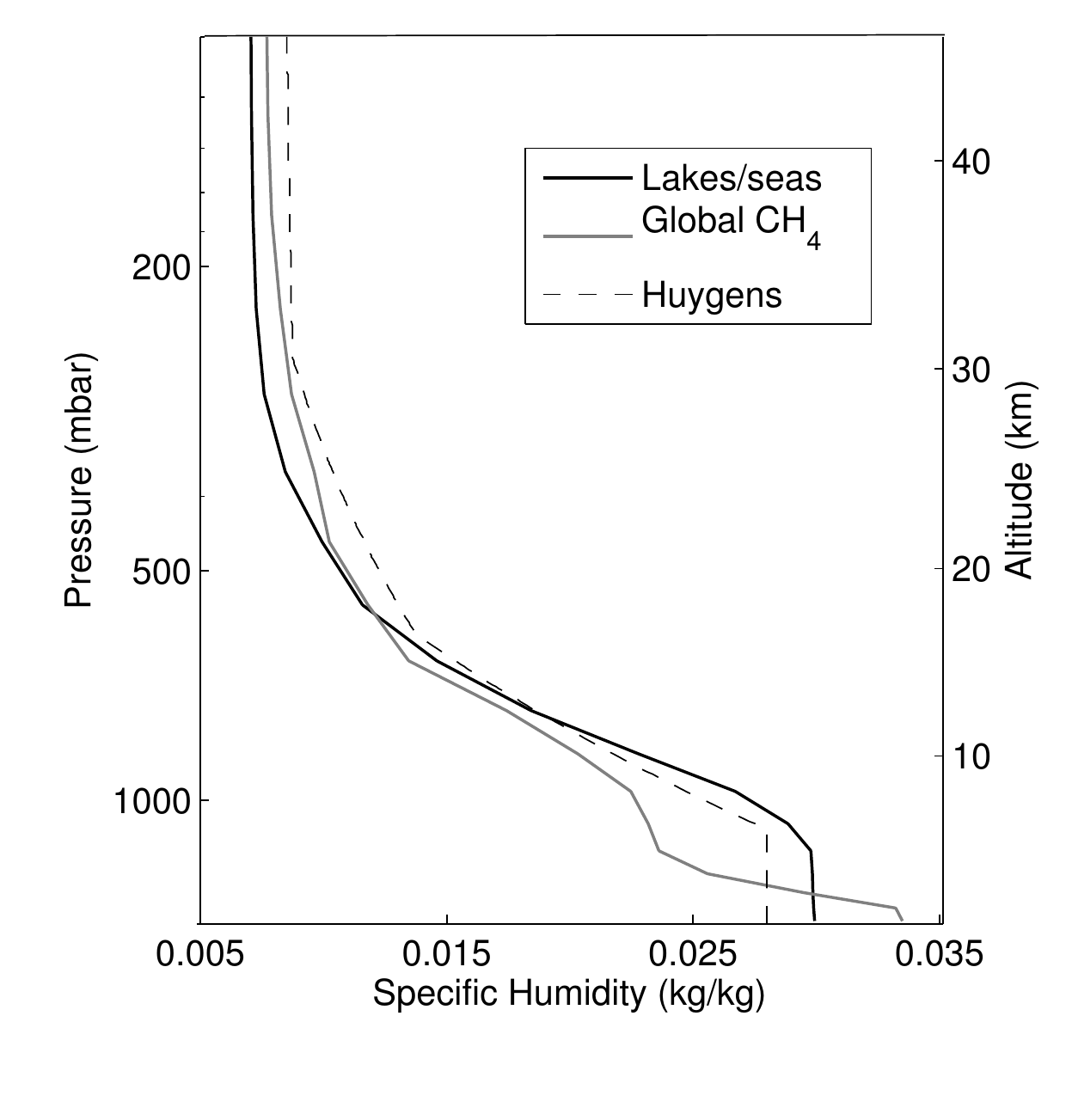}
	\caption[Methane specific humidity profile]{Modeled (zonally averaged) and observed methane specific humidity near the equator around the time of the Huygens probe descent. An approximate altitude scale is included for reference.
	\label{Fig:ch4_profile}}
	\end{center}
\end{figure}

Modeled and observed \citep{Niemann05} equatorial tropospheric methane profiles are shown in Fig.~\ref{Fig:ch4_profile}. In both simulations, the specific humidity at the surface is high, compared to that measured, but the global methane simulation overestimates it significantly more. Additionally, the lakes/seas simulation produces a nearly-constant specific humidity at pressures above $\sim$1100 mbar, as observed, whereas the profile in the other case is distinctly different, with the specific humidity increasing almost to the the surface. This is a consequence of the availability of moisture from the surface, which indicates that, within our assumptions, the observed methane profile at low latitudes is consistent with a dry surface.

\par
The turn-over in the lakes/seas simulation equatorial methane profile starting around $\sim$1160 mbar also corresponds to a transitional region in the global methane simulation profile. This is because that level in the atmosphere marks the temperature (87 K) chosen as the transition between the methane-nitrogen liquid and methane ice in the computation of saturation vapor pressure. This temperature is high compared to the standard assumption that the condensate is liquid down to around 80 K, but agrees quite well with the altitude where the relative humidity of CH$_4$-N$_2$ liquid, calculated from the observed methane profile, stops increasing linearly with altitude \citep{Tokano06}, and may mark the beginning of the transition from a liquid mixture to a pure methane ice. Regardless, the simulated increase of specific humidity toward the surface would occur, with slightly different values, in either case, as long as a source of methane were present on the surface; this behavior was not observed and is not produced in the lakes/seas simulation.
\par
In all cases, the simulated methane profile above the tropopause (not shown) increases slightly to a specific humidity of about 0.01 at 10 mbar, and is constant at lower pressures. This is a slight over-estimate compared to the observations, and the increase represents a too-large flux of methane between troposphere and stratosphere, despite the cold trap of the tropopause; the mechanism for this is not clear, but is potentially related to the lack of a sink for methane at the model top. Nevertheless, this discrepancy has a negligible effect on the methane cycle of the lower troposphere and surface.
\par
\begin{figure}[h!]
	\begin{center}
	\includegraphics[width=0.8\textwidth]{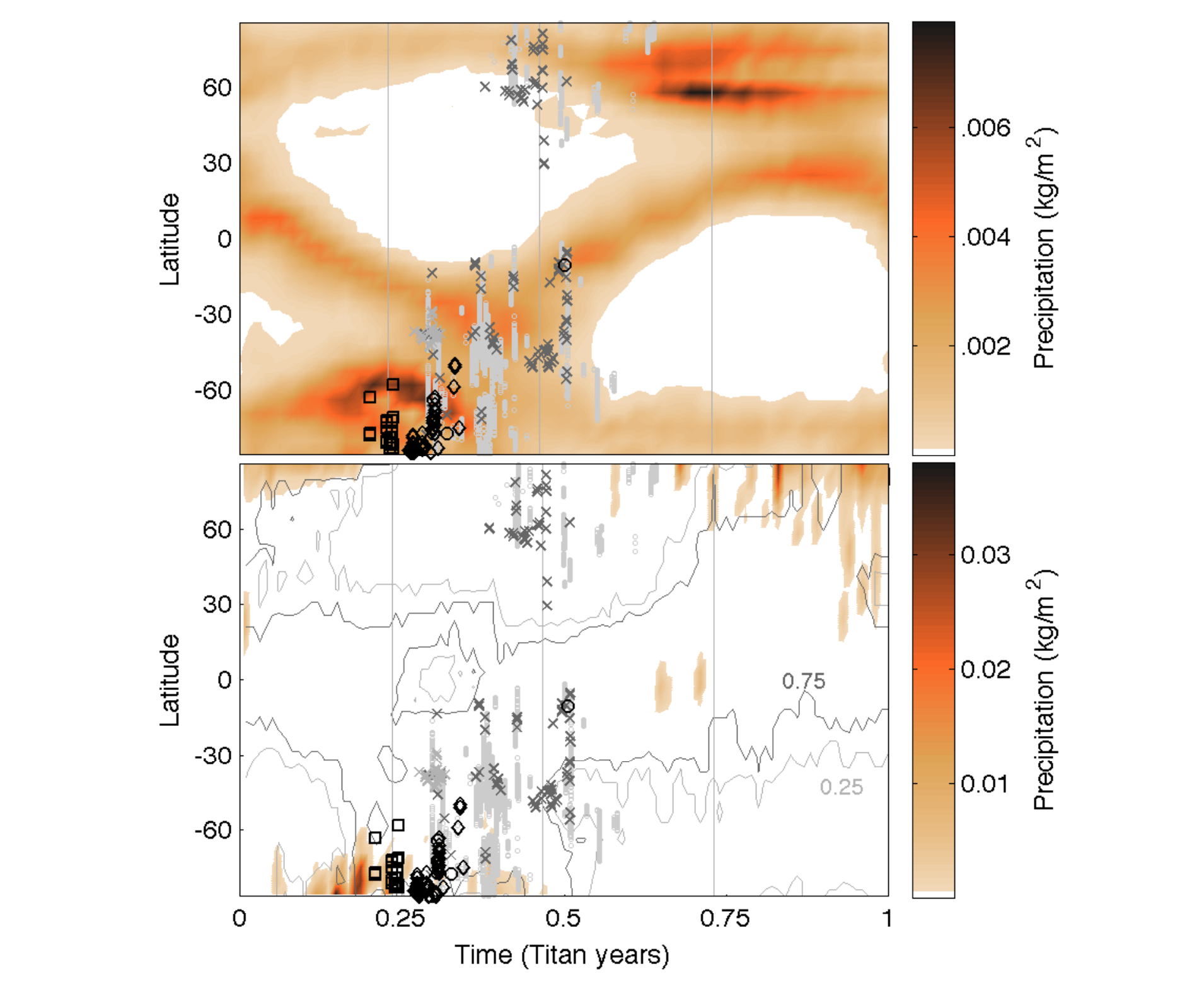}
	\caption[Precipitation maps]{Top: Precipitation (kg/m$^2$) distribution for the L32 global surface reservoir simulation, averaged over five Titan years, in color contours. Bottom: The same for the L32 lakes/seas simulation in color. Additional gray contours show the distribution of the frequency of large-scale condensation in the troposphere (see text); the light and dark contour lines are 0.25 and 0.75, respectively. Observations of clouds are shown in both panels for comparison: Black squares are from \citet{Bouchez05}; black diamonds are from \citet{Schaller06b}; small light gray circles are observations from VIMS \citep[S. Rodriguez, personal communication;][]{Rodriguez09,Rodriguez11}; black circles are clouds labeled ``convective" from \citet{Turtle11a}; Light gray x's are from \citet{Roe05b}; and darker gray x's are clouds labeled other than ``convective" from \citet{Turtle11a}. Note the different scales for the color contours. Vertical lines indicate the timing of solstices and northern vernal equinox.
	\label{Fig:precip}}
	\end{center}
\end{figure}

Distributions of precipitation versus time are shown in Fig.~\ref{Fig:precip}, with some cloud observations overlain for comparison (with the assumption that model precipitation can be used as a proxy for clouds). Most of this is moist convective precipitation that immediately reaches the surface. Relatively light but sustained precipitation is prevalent in the global surface reservoir simulation, and a clear relationship exists between the location of low and mid-latitude precipitation and that of seasonally-controlled upwelling, in agreement with previous models \citep{Mitchell06,Mitchell11}. Also present is summertime polar precipitation that decreases but does not cease in other seasons. Though this precipitation distribution appears to agree with the majority of cloud observations, it also implies nearly-permanent cloud cover and continued activity at the south pole during and after equinox, neither of which are observed. In addition, the observed preference for clouds around 40$^{\circ}$ is only satisfied because of the full-hemisphere coverage of precipitation for half of Titan's year; the particular latitudes are not actually preferred.

\par
In the lakes/seas simulation, precipitation is by comparison much more sparse but at times up to an order of magnitude more intense. Summertime polar precipitation is robust, as are occasional low-latitude outbursts. The latter are in agreement with some of the data, which correspond to large (observed by ground-based telescopes) events and clouds labeled ``convective" by \citet{Turtle11a}. Indeed, pauses in activity occur after precipitation outbursts, in agreement with the suggestion that atmospheric depletion inhibits subsequent convection \citep{Schaller06b}. However, this precipitation distribution does not match well with other observations of clouds, particularly in the mid-latitudes. Those clouds display characteristics consistent with convective systems \citep{Griffith05}, but also tend to exhibit different, elongated morphologies compared to the polar clouds that were prevalent shortly after solstice \citep{Turtle11a}. 
\par
The bottom panel of Fig.~\ref{Fig:precip} also shows the distribution of the frequency of large-scale condensation in the troposphere between the surface and approximately 500 mbar (gray contours). While the vast majority of this condensation does not produce precipitation that reaches the ground, its distribution is similar to that of precipitation in the global-methane simulation. Mid-latitude cloud observations fall within regions where condensation occurs fairly frequently. Thus, this large-scale condensation provides a possible explanation for mid-latitude clouds, as well as for the optically-thin stratiform clouds tentatively detected from \textit{in situ} data \citep{Tokano06}. Note that at pressures lower than 500 mbar (not shown), light large-scale condensation is frequent during polar winter as a result of decreasing temperatures at the tropopause. This is a different mechanism associated with higher-altitude, non-convective cloud decks. Similar cloud decks have been observed over the winter polar tropopause \citep{Griffith06,LeMouelic12}; however, these form by condensation of downwelling species from the stratosphere and are composed primarily of ethane, so are not captured by the model.
\par
\begin{figure}[h!]
	\begin{center}
	\begin{tabular}{c}
		\includegraphics[width=0.7\textwidth]{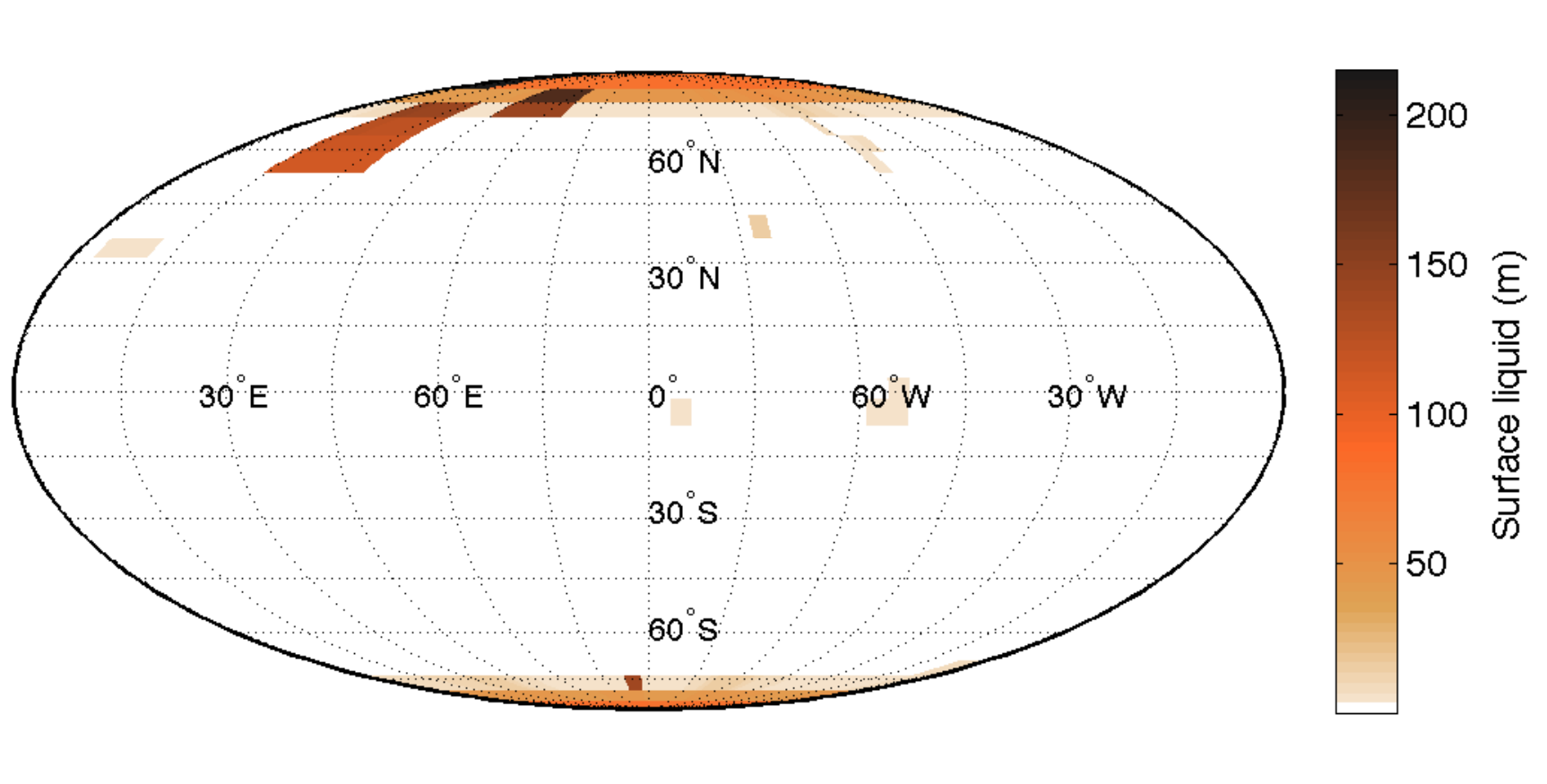} \\
		\includegraphics[width=0.7\textwidth]{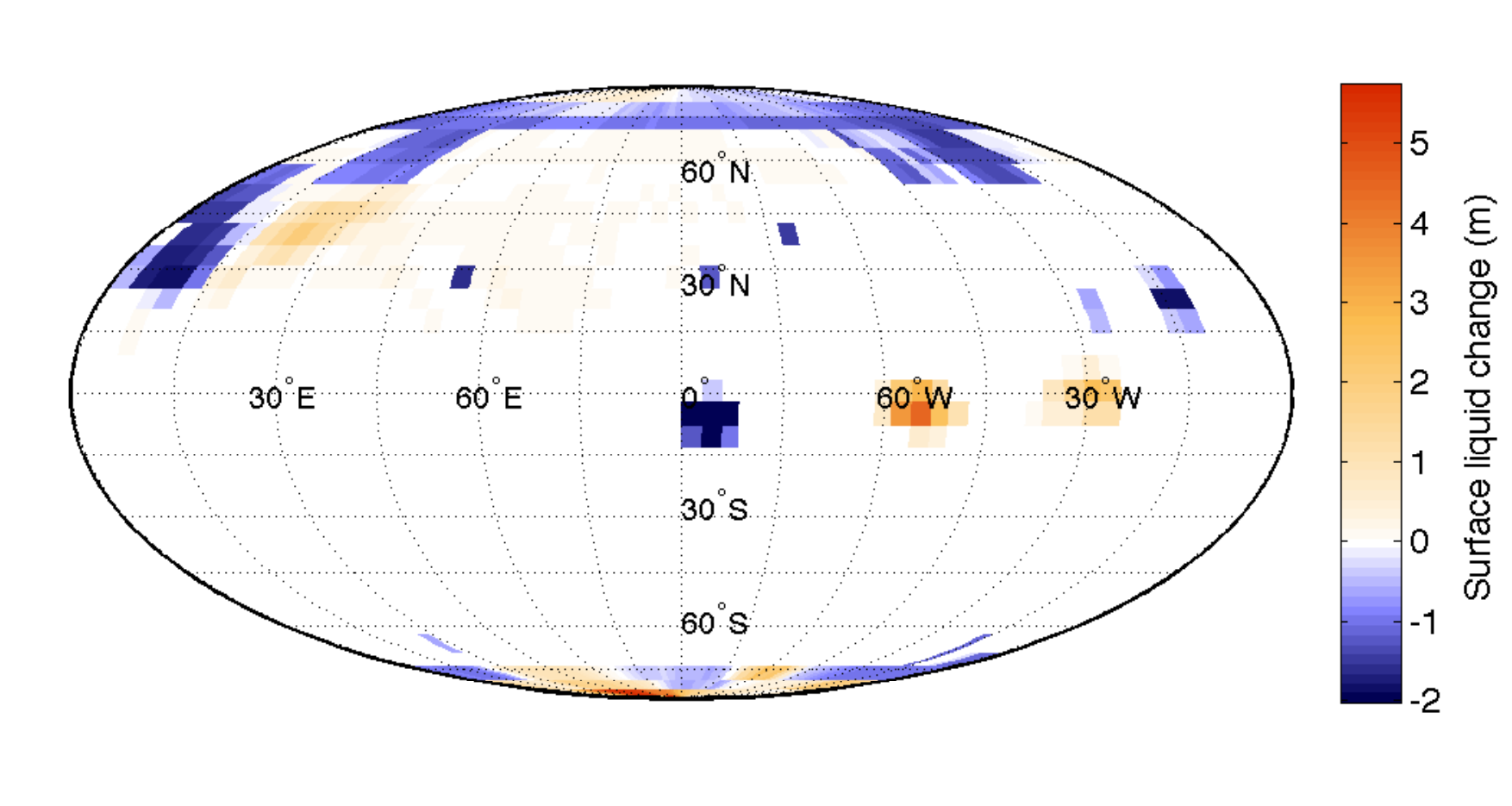}
	\end{tabular}
	\caption[Surface liquid maps]{Top: Map of surface liquid (m) at northern summer solstice in the last year of L32 lakes/seas simulation. The four darkest features are the initialized lakes/seas, which have remained above 100 m in depth throughout the simulation. Bottom: The net change in surface liquid (m) over the course of the last year of simulation, between northern autumnal equinoxes. Note that several of the features in the bottom map are not visible in the top, indicating their ephemeral nature. Both maps are Mollweid equal area projections.
	\label{Fig:qsurf_map}}
	\end{center}
\end{figure}

The surface liquid coverage near the end of the lakes/seas simulation is shown in Fig.~\ref{Fig:qsurf_map}, along with the net change during the last year of simulation. Methane accumulates at both poles in agreement with previous results \citep{Schneider12}, with a slight enhancement in coverage visible in the north. Surface methane in the model is highly unstable at latitudes $<$50$^{\circ}$, and is quickly transported poleward by the atmosphere, as in previous models \citep{Rannou06,Mitchell06,Mitchell08,Mitchell09,Mitchell12,Schneider12}. (Note that there is no build-up of surface methane at mid-latitudes as the equatorial surface dries, as in the results of \citet{Mitchell08}.) The simulated atmosphere holds roughly 5~m of precipitable methane, which agrees with previous models and observations \citep{Schneider12,Tokano06}. Patchy surface liquid at low latitudes, which coincides with the detection of equatorial lakes \citep{Griffith12}, is associated with bursts of precipitation there. But, as shown by the bottom panel of Fig.~\ref{Fig:qsurf_map}, these features are shallow and ephemeral. Interestingly, there is increased activity at mid latitudes in the vicinity of the large northern seas, but again no significant buildup remains. 

\par
While seasonal activity is clearly an important mechanism in the development of precipitation (Fig.~\ref{Fig:precip}), the availability of surface methane appears to be a prerequisite to rain, with the notable exception of the occasional low-latitude outbursts reminiscent of those observed around equinox \citep{Schaller09,Turtle11b}, and previously explained as due to 3D wave activity \citep{Mitchell11}. Consistent precipitation at high latitudes is clearly linked to the availability of methane at the surface, as well as the insolation. Without invoking a physically implausible fast sub-surface transport of liquid \citep[i.e.,][]{Schneider12}, only simulations with an inexhaustible, global surface reservoir produce any significant precipitation at summer mid-latitudes. Also considering the dearth of observed lakes/seas away from polar regions and the observed prevalence of 40$^{\circ}$S clouds before, during, and after equinox, this suggests that mid-latitude cloud activity is either non-convective and non-precipitating, or caused by a mechanism not currently included in the model that is only somewhat related to the changing seasons. Some possible such mechanisms might be topographical forcing (i.e., via orographic gravity waves), a sub-surface source of methane (possibly cryovolcanism or seepage from a methane table), or another non-convective/non-precipitating form of cloud formation. It is worth noting that early studies of these clouds suggested a longitudinal as well as latitudinal dependence \citep{Porco05,Roe05b}, though later analyses disputed this \citep{Griffith05,Brown10,Rodriguez11}.
\par
\begin{figure}[h!]
	\begin{center}
	\includegraphics[width=0.7\textwidth]{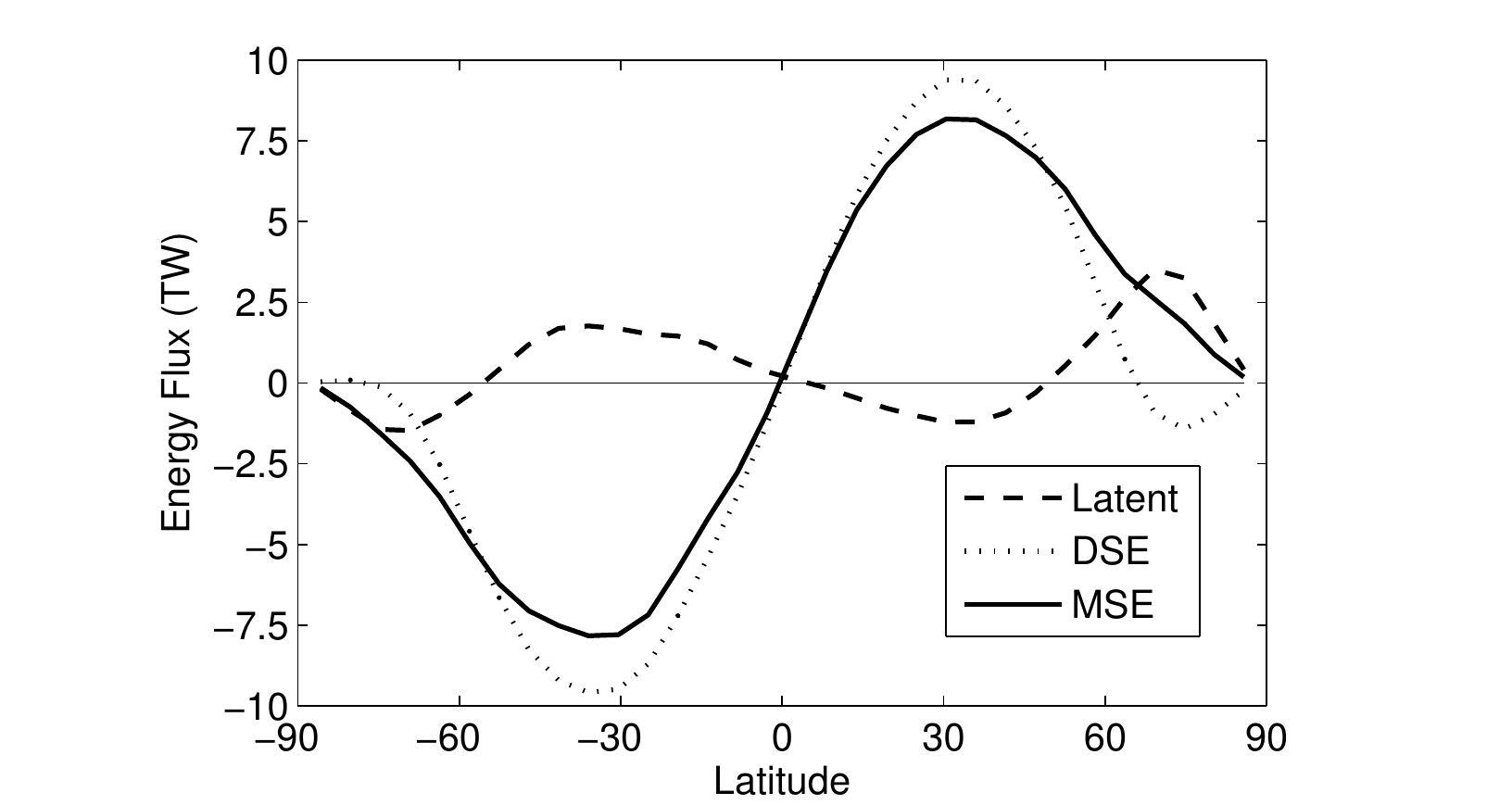}
	\caption[Atmospheric energy fluxes]{Annual-mean atmospheric energy fluxes from the L32 lakes/seas simulation. ``DSE" and ``MSE" stand for dry static energy and moist static energy, respectively. Positive values indicate northward flux.
	\label{Fig:energy_fluxes}}
	\end{center}
\end{figure}

The mean meridional energy transport by the atmosphere can be examined via fluxes of moist static energy (MSE), which is the sum of dry static energy (DSE; internal plus potential energy in an air parcel) and latent energy due to moisture. Fig.~\ref{Fig:energy_fluxes} shows the annual-mean fluxes of vertically integrated moist static, dry static, and latent energies as a function of latitude for the lakes/seas simulation. The MSE flux is dominated by the flux of DSE at all but the highest latitudes, and is divergent at the equator. On the other hand, latent energy flux, which dominates at high latitudes and is asymmetric (with a net northward value), is convergent at the equator; latent energy flows opposite to the MSE flux, dominantly transported by near-surface air from winter to summer hemispheres. The divergence of the latent energy flux at mid-latitudes illustrates the atmosphere's ability to transport methane away from these regions.
\par
\citet{Griffith14} suggest that cold-trapped polar methane \citep{Schneider12} may explain the observed equatorial humidity ($\sim$50$\%$), since the observed equator to pole surface temperature gradient would imply $\sim$85$\%$ polar humidity with the same methane content. This would agree with estimates, based on energy arguments, of low advective transport of methane \citep{Griffith08}. However, \citet{Mitchell12} showed that Titan's constant outgoing longwave radiation with latitude is evidence of transport by the atmosphere, a large portion of which is done by latent energy fluxes. Our results provide more realism by eliminating the global methane source assumed in \citet{Mitchell12}, but nevertheless show significant transport of methane, via which polar moisture humidifies the equatorial atmosphere in agreement with \citet{Griffith14}.

\section{Discussion}\label{Sec:discussion}
The simulations presented in this paper represent an effort to model the circulation and climate of Titan's atmosphere realistically, eliminating several significant simplifications from past models and succeeding in reproducing many important aspects of the atmosphere. Nevertheless, the model still necessarily employs a variety of simplifications that are presently discussed.
\par
The relatively low wind speeds in the model's high stratosphere are not particularly surprising given the variety of factors that are probably contributing to inhibiting the magnitude of superrotation. These include (but are probably not limited to): The low model top, with a sponge layer to prevent spurious wave reflections, which artificially damps the winds of the top layer and probably significantly affects the top several layers' circulation; the lack of haze-dynamics coupling, which is shown to depress wind-speeds at high altitudes \citep[as in][]{Rannou02,Rannou04} and should increase latitudinal temperature contrasts \citep{Rannou04,Crespin08}; the assumption of hydrostatic balance that ignores the effect of the wind-induced equatorial bulge \citep{Tokano13}; and the assumption of horizontally-homogeneous, dynamically uncoupled stratospheric trace gases that are radiatively active. Indeed, the zonal wind results at pressures higher than $\sim$1 mbar are excellent, and the model's capability to build up atmospheric angular momentum is quite promising. Further upgrades to alleviate the above restrictions are the subject of future work.
\par
Several aspects of the methane cycle should also be considered simplifications that warrant further development. Importantly, only methane is included as a tracer, so the depressing effect of ethane on evaporation rates, for instance, is neglected. Related to this is the simplified calculation of vapor pressure, in which the effects of dissolved nitrogen are only roughly accounted for. The inclusion of additional tracers for the methane cycle, such as methane ice particles, and coupling to the radiative transfer, may also improve the accuracy and fidelity of simulated tropospheric clouds.
\par
Furthermore, no microphysical considerations were included in the simulation of cloud formation or precipitation. The moist convective parameterization provides an improvement over exclusively including large-scale condensation, which, particularly at this resolution, is probably inaccurate. However, quasi-equilibrium convection still produces precipitation that is fairly sporadic, and is only an idealized representation of the process. Given the discrepancy between the distribution of precipitation and observed tropospheric clouds, and considering that several mechanisms may be involved in cloud formation, a more detailed condensates scheme as implemented in other models \citep{Rannou06,Burgalat14}, may also be justified.
\par
The surface model, while showing that an infinite source of methane constantly available to the atmosphere is inconsistent with the observations, could also benefit from a considerable increase in complexity. An obvious improvement is the inclusion of topography, which would allow for testing of the importance of orographic forcing on cloud formation \citep{Roe05b,Porco05}. Including surface runoff might also prove useful, for example in simulating and predicting the locations of small lakes and lake-effect or ``marine" clouds \citep{Brown02}, though perhaps only at higher resolutions. Surface thermal properties, as well as albedos, should also be allowed to vary.

\section{Conclusions}\label{Sec:conclusions}
We have presented results from simulations using TAM, a new, fully three-dimensional GCM of Titan's atmosphere with realistic radiative transfer, as well as moist processes and a surface liquid model. Benchmarked against the available observations, our work demonstrates that two of the most important factors for simulating the key aspects of Titan's atmosphere are accurate radiative transfer and a dynamical core numerically capable of developing superrotation.
\par
Several aspects relating to the state of the surface-atmosphere system are elucidated through the simulations:

\begin{itemize}
\item{The vertical temperature profile through the stratopause, both at the equator and at the poles, is reproduced satisfactorily without the need to invoke complex interactions between haze, trace gases, and dynamics (though these may further improve the results \citep{Rannou02,Rannou04,Crespin08}). CIRS measurements are particularly well reproduced. The polar structure observed in radio occultations \citep{Schinder12} is similar to what is produced by a temperature oscillation that originates in the high stratosphere at the onset of polar night and propagates downward, though the simulated timing precedes the observations.}
\item{No additional physics or numerical techniques are necessary to achieve proper atmospheric superrotation. The zonal wind profile through the lower stratosphere agrees well with observations, with the exception of the observed minimum around 75 km altitude; this structure may be related to both the polar temperature oscillation and/or the seasonal variation of haze and its radiative effects.}
\item{Surface turbulent fluxes respond to the surface insolation, and as a result their maxima occur over summer mid-latitudes, not the polar regions, counter to what has been previously suggested.}
\item{Surface liquids quickly migrate polewards from lower latitudes, in agreement with prior studies \citep{Rannou06,Mitchell06,Mitchell08,Schneider12,Lora14}; they are unstable equatorward of approximately 60$^{\circ}$ on timescales of order one Titan year. Both the surface temperature and the vertical distribution of humidity at low latitudes are consistent with this distribution of surface liquids. Atmospheric energy fluxes are also consistent with this picture, and also indicate a preference toward northward methane transport in the current epoch, as in \citet{Schneider12}.}
\item{Summer mid-latitude clouds, however, are difficult to explain with the simulated precipitation distribution, except when global surface liquid is imposed, and then not very satisfactorily. We suggest that these clouds, as observed on Titan, are either non-precipitating, or are related to a process not currently captured in this GCM.}
\end{itemize}

Titan's atmosphere and surface represent a complex system in which a variety of factors communicate. With this work, it is evident that realistic interactions between physical processes like radiation and variable surface moisture---seldom considered together in previous models---are critical to properly simulating the climate of this unique world. Future studies, in particular those involving the methane cycle, must include proper radiative transfer and move beyond the assumption of a global ocean on Titan.

\section*{Acknowledgements}
The authors acknowledge support from NASA Earth and Space Science Fellowship\\NNX12AN79H, and the Cassini project. Simulations were carried out with an allocation of computing time on the High Performance Computing systems at the University of Arizona. The authors would also like to thank S. Rodriguez and an anonymous reviewer for detailed comments to improve the manuscript, and S. Rodriguez for providing the VIMS cloud observations.

\end{document}